\documentclass[12pt,a4paper,dvips]{article}

\usepackage{a4p}
\usepackage{cite,mcite}
\usepackage{graphicx}
\usepackage{physics}
\usepackage{wasysym}

\usepackage{l3_titlePH,ifthen}

\journalname{Phys. Lett. B}

\date{April 4, 2005}

\preprint{2005-013}

\newlength{\capindent}
\newlength{\capwidth}
\newlength{\figwidth}
\newcommand{\icaption}[2][!*!,!]{\hspace*{\capindent}%
  \begin{minipage}{\capwidth}
    \ifthenelse{\equal{#1}{!*!,!}}%
      {\caption{#2}}%
      {\caption[#1]{#2}}
  \end{minipage}}
\newcommand{\Zg}{\ensuremath{\rm Z\gamma^{*}}}

\newcommand{\qqll}{\ensuremath{\rm q \bar q\ell^+\ell^- }}
\newcommand{\llnn}{\ensuremath{\rm \ell^+\ell^- \nu\bar\nu}}

\newcommand{\llll}{\ensuremath{\rm \ell^+\ell^-\ell^{\prime +}\ell^{\prime -} }}
\newcommand{\qqnn}{\ensuremath{\rm q\bar q\nu\bar\nu}}

\newcommand{\qqee}{\ensuremath{\rm q\bar q e^+ e^-}}
\newcommand{\qqmm}{\ensuremath{\rm q\bar q \mu^+ \mu^-}}

\begin{document}
\begin{titlepage}
\title{Neutral-Current Four-Fermion Production \\
in \boldmath{\epem} Interactions at LEP}
\author{The L3 Collaboration}
%
%
\begin{abstract}

Neutral-current four-fermion production, $\epem \ra f\bar{f}f'\bar{f}'$,
is studied in 0.7~fb$^{-1}$ of data collected with the L3 detector at
LEP at centre-of-mass energies $\sqrt{s} = 183 - 209\GeV$. Four final
states are considered: \qqnn{}, \qqll{},  \llll{} and \llnn{}, where
$\ell$ denotes either an electron or a muon. Their cross sections are
measured and found to agree with the Standard Model predictions. In
addition, the $\epem \ra{\rm Z} \gamma^* \ra f\bar{f}f'\bar{f}'$
process is studied and its total cross section at the average
centre-of-mass energy $\langle \sqrt{s} \rangle = 196.6\GeV$ is found
to be $0.29\pm 0.05 \pm 0.03~\rm pb$, where the first uncertainty is
statistical and the second systematic, in agreement with the Standard
Model prediction of 0.22~pb.  Finally, the mass spectra of the \qqll{}
final states are analysed to search for the possible production of a
new neutral heavy particle, for which no evidence is found.

\end{abstract}

\submitted

\end{titlepage}

%
\section{Introduction}
%

The centre-of-mass energy, $\sqrt{s}$, of the LEP $\epem$ collider reached
$209 \GeV$,  allowing the study of four-fermion production mediated by
the exchange of real and virtual gauge-bosons.  Four-fermion events
are classified as charged-current processes or neutral-current
processes~\cite{CCNC,LEP2YR}. The former proceed through the exchange
of W bosons, while the latter comprise the exchange of both Z bosons
and off-mass-shell photons.

\begin{figure}[p]
  \begin{center}
    \begin{tabular}{cc}
      \mbox{\includegraphics*[width=0.45\textwidth]{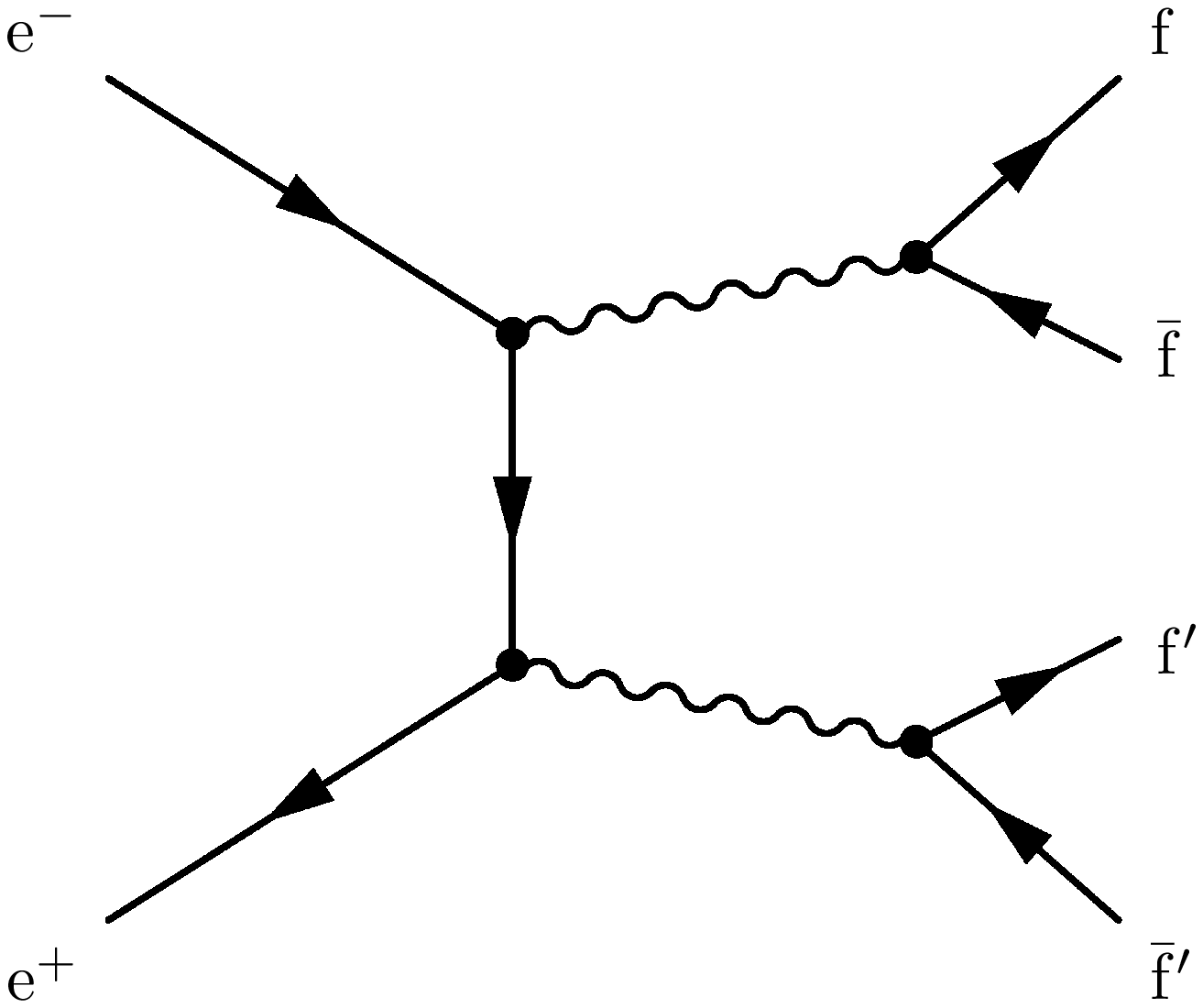}} \hspace{3ex}&
      \mbox{\includegraphics*[width=0.45\textwidth]{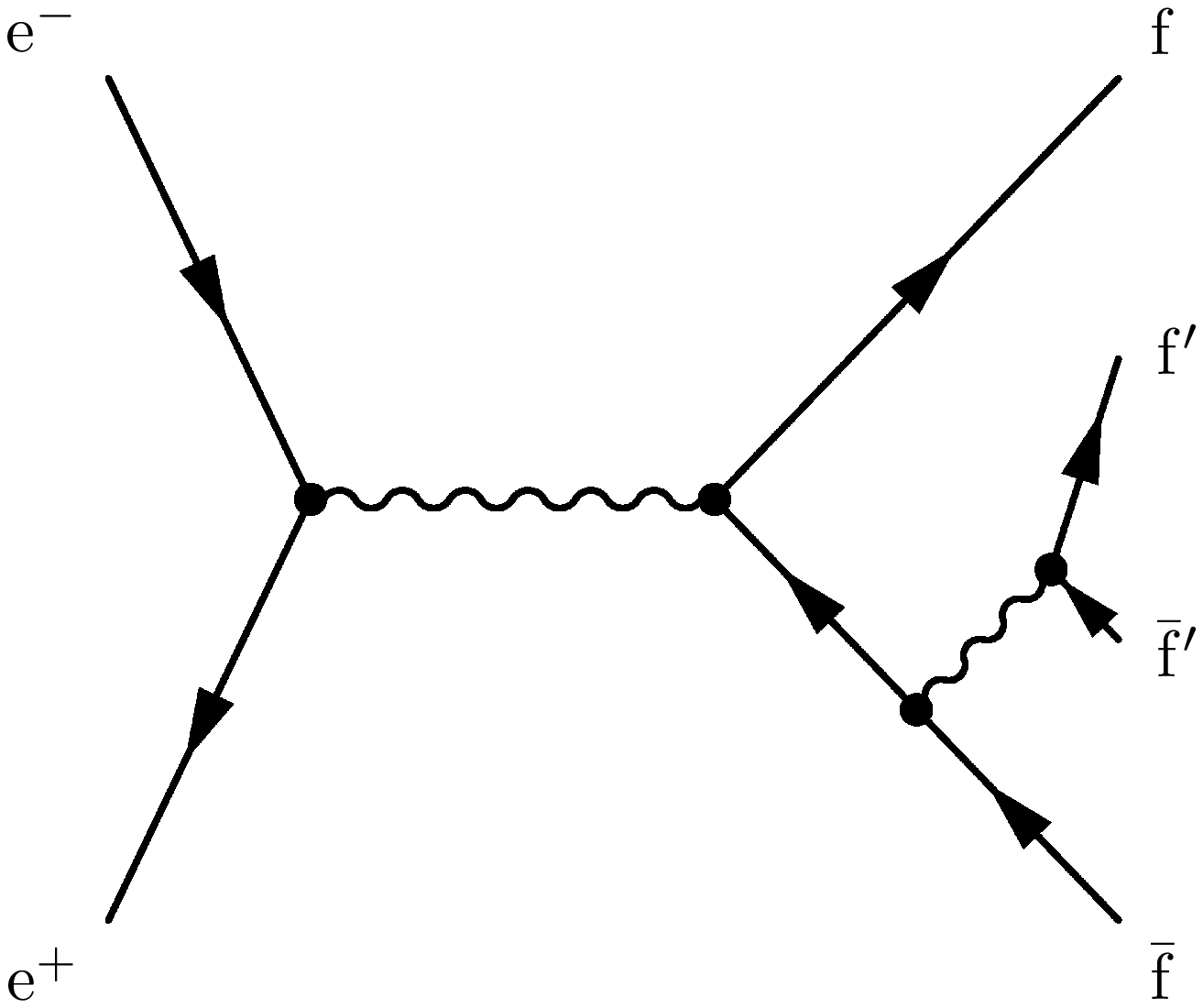}}\\
      \mbox{Conversion} \hspace{3ex}&
      \mbox{Annihilation} \vspace{7ex}\\
      \mbox{\includegraphics*[width=0.45\textwidth]{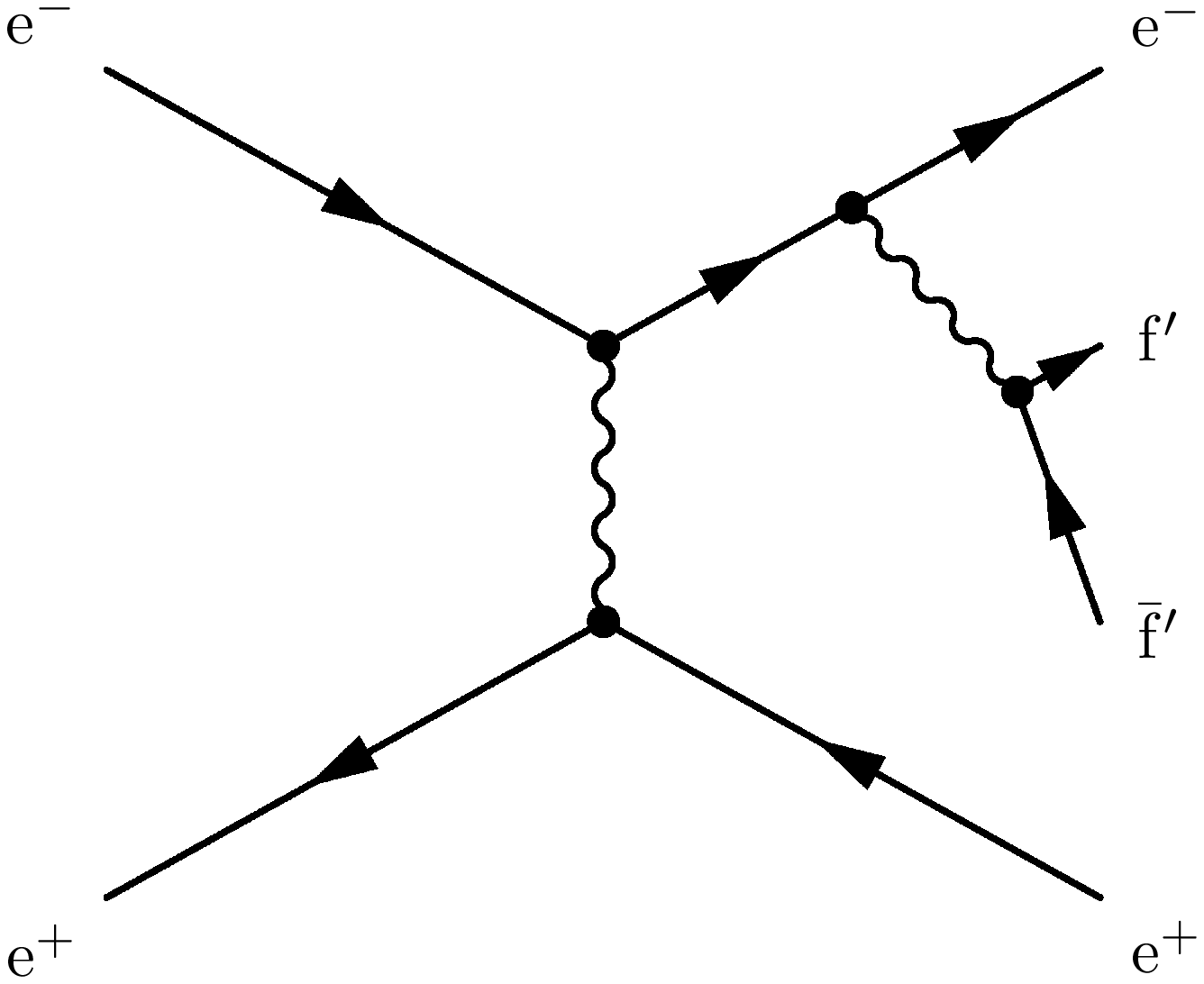}} \hspace{3ex}&
      \mbox{\includegraphics*[width=0.45\textwidth]{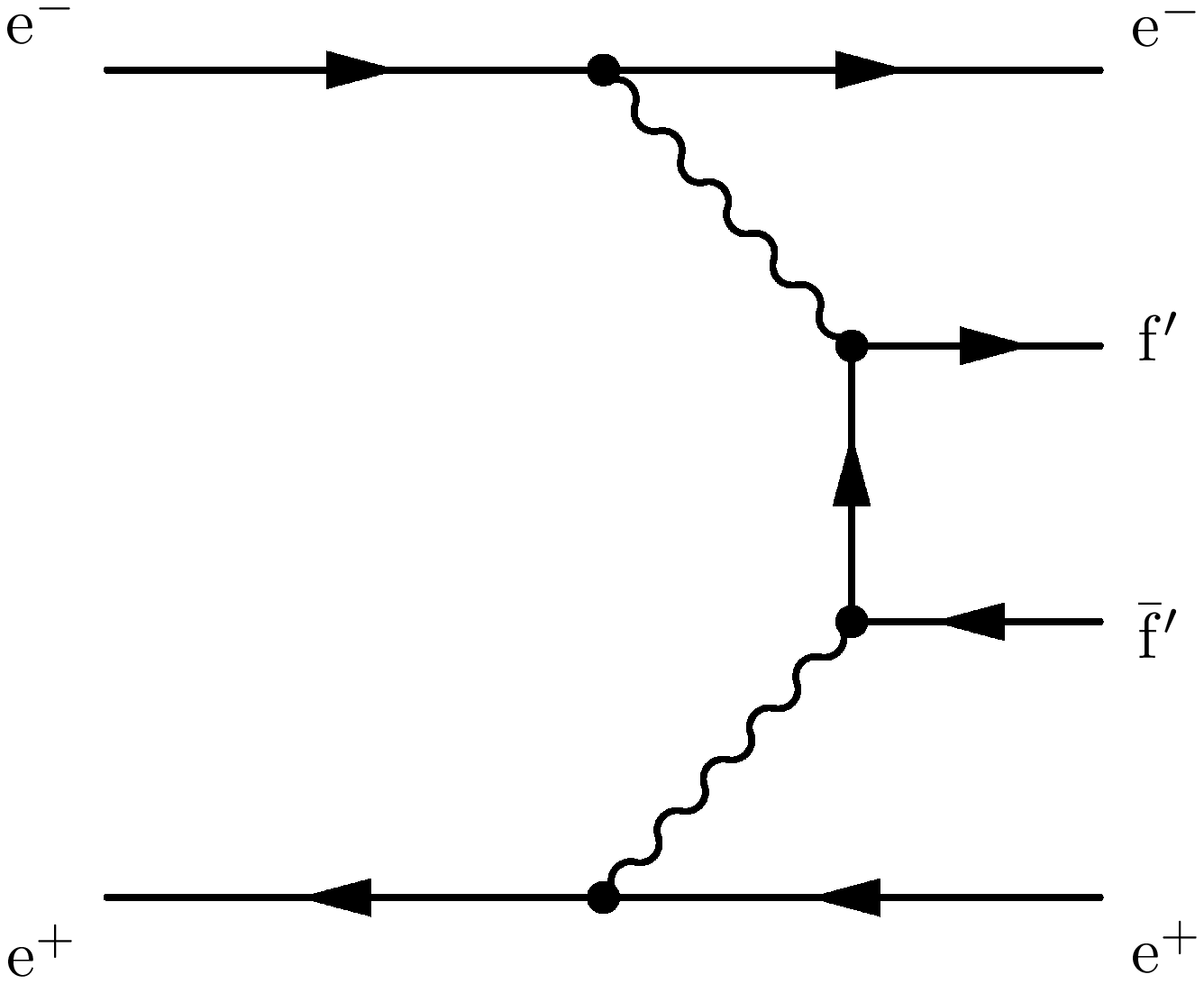}} \\
      \mbox{Bremsstrahlung} \hspace{3ex}&
      \mbox{Multiperipheral} \\
    \end{tabular}
    \caption{Dominant Feynman diagrams contributing to neutral-current
    four-fermion production. The wavy lines represent a Z boson or an
    off-mass-shell photon.}
    \label{fig:1}
  \end{center}
\end{figure}

The L3 collaboration has investigated both charged-current processes,
in particular W-boson pair production~\cite{wPairs} and single W-boson
production~\cite{singleW}, and neutral-current processes, with Z-boson
pair production~\cite{zPairs} and single Z-boson
production~\cite{singleZ}. Reference~\citen{lepew} provides
a comprehensive bibliography of these studies at LEP. This Letter
extends previous studies, focused on the signature of boson pairs or
a single boson and missing energy, to a general analysis of
neutral-current processes, $\epem\ra f\bar{f}f'\bar{f}'$, with either
high-mass or low-mass fermion pairs. The special case of events with a
Z boson and an off-mass-shell photon, $\epem \ra{\rm Z} \gamma^* \ra
f\bar{f}f'\bar{f}'$, is also considered. Figure~\ref{fig:1} presents
some of the Feynman diagrams responsible for neutral-current
four-fermion production. In the following, four processes with
different combinations of quarks, leptons and neutrinos in the final
state are considered:  $\epem\ra\qqnn$, $\epem\ra\qqll$,
$\epem\ra\llnn$ and $\epem\ra\llll$, where $\ell$ denotes either an
electron\footnote{Throughout this Letter the term ``electron'' stands
for both electrons and positrons.}  or a muon. Events with tau leptons
are considered as background.

High-resolution studies of four-fermion events are also sensitive to
manifestations of new physics, and the mass spectra of events from the
$\epem\ra\qqll$ process are investigated to search for new neutral
heavy particles decaying into hadrons.

L3 results on neutral-current four-fermion production in a smaller
data sample obtained at lower values of $\sqrt{s}$ are discussed in
References~\citen{L3130-172}~and~\citen{L3183}. A study of the
$\epem\ra\qqll$ process by the OPAL collaboration is described in
Reference~\citen{OPALqqll}.

\begin{table}[hb]
  \begin{center}
    \begin{tabular}{|rcl|l|}
      \hline
      \multicolumn{3}{|c|}{Process} & \multicolumn{1}{c|}{Signal definition} \\
      \hline
      \rule{0pt}{12pt} $\epem$&$\ra$&$\qqnn$ & $m_{\rm\qqbar} > 10 \GeV$ \\
      \rule{0pt}{12pt} $\epem$&$\ra$&$\qqll$ & $|\cos\theta_\ell|<0.95$,  $m_{\ell^+\ell^-} > 5 \GeV$,  $m_{\rm\qqbar} > 10 \GeV$ \\
      \rule{0pt}{12pt} $\epem$&$\ra$&$\llnn$ & $|\cos\theta_\ell|<0.95$, $m_{\ell^+\ell^-} > 5 \GeV$,  $m_{\ell\nu} \notin [70\GeV$, $90\GeV$] \\
      \rule{0pt}{12pt} $\epem$&$\ra$&$\llll$ & $|\cos\theta_\ell|<0.95$, $|\cos\theta_{\ell'}|<0.95$,  $m_{\ell^+\ell^-} > 5 \GeV$, 
                                  $m_{\ell'^+\ell'^-} > 5 \GeV$, \\
      \rule{0pt}{12pt} &&&  $m_{\ell^\pm\ell'^\mp} > 5 \GeV$ if $\ell$ and $\ell'$ have same flavour\\
      \hline
      \rule{0pt}{12pt} $\epem$&$\ra$&${\rm Z} \gamma^* \ra f\bar{f}f'\bar{f}'$ & 
                                  $m_{f\bar{f}} \in [m_{\rm Z} -2\Gamma_{\rm Z}, m_{\rm Z} +2\Gamma_{\rm Z}]$ and 
                                  $m_{f'\bar{f'}} \notin [m_{\rm Z} -2\Gamma_{\rm Z}, m_{\rm Z} +2\Gamma_{\rm Z}]$ or\\
      \rule{0pt}{12pt} &&&
                                  $m_{f\bar{f}} \notin [m_{\rm Z} -2\Gamma_{\rm Z}, m_{\rm Z} +2\Gamma_{\rm Z}]$ and
                                  $m_{f'\bar{f'}} \in [m_{\rm Z} -2\Gamma_{\rm Z}, m_{\rm Z} +2\Gamma_{\rm Z}]$ \\
      \hline
    \end{tabular}
    \caption{Signal definition for neutral-current four-fermion final states}
    \label{tab:0}
  \end{center}
\end{table}

The studies of four-fermion final states are restricted to a limited
part of the full phase space as described in Table~\ref{tab:0}.  These
signal-definition criteria have multiple purposes. Cuts on the cosine
of the angle between the leptons and the beam axis,
$|\cos\theta_\ell|$ and $|\cos\theta_{\ell'}|$, restrict the
comparison between data and predictions to regions compatible with the
geometrical coverage of the detector, thus avoiding large
extrapolation factors. Cuts on the masses of the fermion-antifermion
pairs, $m_{\ell^+\ell^-}$, $m_{\ell'^+\ell'^-}$ and $m_{\rm\qqbar}$,
remove contributions of strongly-interacting resonances in the
low-mass regions.  If four same-flavour leptons are produced in the
$\epem \ra\llll$ process, an additional mass cut is applied to account
for all possible lepton combinations.  The $\epem \ra \llnn$ process
is mostly due to charged-current W-boson pair production, studied in
detail elsewhere~\cite{wPairs}. A cut is applied on the
lepton-neutrino mass, $m_{\ell\nu}$, to reduce the contribution of W
bosons and enhance that of neutral-current four-fermion production.

Two additional phase-space criteria, also listed in Table~\ref{tab:0},
are applied to increase the relative contribution from the $\epem
\ra{\rm Z} \gamma^* \ra f\bar{f}f'\bar{f}'$ process. The mass of one
of the fermion-antifermion pairs is required to be in the range
$m_{\rm Z}\pm 2\Gamma_{\rm Z}$, while the mass of the other pair is
required to be outside this range, where $m_{\rm
Z}=91.19\GeV$~\cite{pdg} and $\Gamma_{\rm Z}=2.49\GeV$~\cite{pdg} are
the Z-boson mass and width, respectively.

%
\section{Data and Monte Carlo samples}
%

The full data-sample collected at high centre-of-mass energies with
the L3
detector~\cite{l3} is
investigated. It amounts to 673.4~pb$^{-1}$ of integrated luminosity
for $\sqrt{s}=182.7-209.2\GeV$, with a luminosity-weighted average
centre-of-mass energy $\langle\sqrt{s}\rangle=196.6\GeV$. The data
were collected around eight average $\sqrt{s}$ values, listed in
Table~\ref{tab:1} together with the corresponding integrated
luminosities. The 55.4~pb$^{-1}$ of data collected at
$\sqrt{s}=182.7\GeV$, already discussed in Reference~\citen{L3183},
are re-analysed within the signal definitions discussed above.

In order to optimise the event selection and calculate the signal
efficiencies, four-fermion events are generated with the
EXCALIBUR~\cite{EXCALIBUR} Monte Carlo program in a phase space larger
than that of the signal definition criteria of
Table~\ref{tab:0}. These cuts are applied on generated quantities and
the selected events are considered as signal, while the remaining ones are
treated as background.  EXCALIBUR is also used to model four-fermion
background from the $\epem\ra\tau^+\tau^- f \bar{f}$ and $\epem\ra\rm
W e \nu$ processes. Additional four-fermion background is due to
W-boson pair production and subsequent decay into fully-hadronic or semi-leptonic
final states. This process is modelled with the KORALW~\cite{KORALW}
Monte Carlo program.  The background from fermion pair-production,
$\epem\ra\qqbar$, $\epem\ra\tau^+\tau^-$ and $\epem\ra\mu^+\mu^-$, is
described by KK2f~\cite{KK2f}. Bhabha scattering is described with
BHAGENE~\cite{BHAGENE} and BHWIDE~\cite{BHWIDE}. The
$\epem\ra\epem\gamma$ process with high transverse-momentum photons
and low polar-angle electrons is simulated with
TEEGG~\cite{TEEGG}. Events with multiple hard-photon production are
generated with GGG~\cite{GGG}. Hadron and lepton production in
two-photon collisions are modelled with PHOJET~\cite{PHOJET} and
DIAG36~\cite{DIAG36}, respectively.

The L3 detector response is simulated using the GEANT~\cite{geant}
program which takes into account the effects of energy loss, multiple
scattering and showering in the detector. GHEISHA~\cite{gheisha} is
used for the simulation of hadronic interactions. Time-dependent
detector efficiencies, as monitored during the data-taking period, are
included in the simulations.

The signal cross sections are calculated with the GRC4F~\cite{grace}
Monte Carlo program which, unlike EXCALIBUR, includes fermion
masses. About twenty thousand events are generated at each value of
$\sqrt{s}$ for each possible flavour combination. The numbers of
events satisfying the criteria in Table~\ref{tab:0} and their weights
are then used to calculate the signal cross sections, listed in
Table~\ref{tab:2}. The cross sections for the $\epem \ra{\rm Z}
\gamma^* \ra f\bar{f}f'\bar{f}'$ processes are extracted for each
final state by applying the additional cuts in Table~\ref{tab:0}, with
the results listed in Table~\ref{tab:4}.

Small differences between the GRC4F and the EXCALIBUR modelling of the 
$\epem \ra f\bar{f}f'\bar{f}'$ process have a negligible impact on the
measurements described in the following.

%
\section{Event selection}
%

The event selection~\cite{miro-thesis+gagan-thesis} is similar to
that devised for the study of Z-boson pair
production~\cite{zPairs}. Events which contain electrons, muons or
hadronic jets are selected and these objects are then combined to
construct kinematic variables to isolate the neutral-current
four-fermion signal from the two-fermion, four-fermion and two-photon
backgrounds.

Electrons are identified by requiring a well-isolated electromagnetic
cluster in the electromagnetic calorimeter with an associated track in
the tracking chamber. To increase efficiency, this track-matching
requirement is relaxed by some selections.

Muons are reconstructed from correlated tracks in the muon
spectrometer and the central tracker which are in time with the beam
crossing. Calorimetric clusters compatible with a minimum ionising
particle with an associated track in the central tracker are also
accepted.

Quark fragmentation and hadronisation yield a high multiplicity of
calorimetric clusters and charged tracks. These are grouped into jets by
means of the DURHAM algorithm~\cite{durham}.

Fermion-antifermion pairs can originate from a Z boson or, for charged
fermions, from an off-mass-shell photon. In the first case, the pair
is characterised by high mass, while in the second case it has most
likely a low mass. Appropriate selections for these two cases are
implemented.

Typical selection variables are: the visible energy of the event,
$E_{\rm vis}$; the transverse, $p_{\rm t}$, and longitudinal, $p_\|$,
components of the vectorial sum of the momenta of all objects in the
event; the missing momentum, $p_{\rm mis}$, and the angles between two
jets or two leptons in space, $\Delta\psi$, or in the plane transverse
to the beam axis, $\Delta\phi$.

The data sample spans a $\sqrt{s}$ range of about $25 \GeV$,
which results in appreciable differences in the kinematics of the
signals. The selection criteria are optimised to reflect these
differences and change over the $\sqrt{s}$ range.

Some aspects of the different selections are described in the
following, while their yields are summarised in Table~\ref{tab:2}.

\subsection{The \boldmath{$\qqnn$} channel}

The signature of the $\epem\ra\qqnn$ process is two hadronic jets and
missing energy mostly due to the production of a Z boson decaying into
neutrinos. The most important background is W-boson pair production.

High-multiplicity events are selected and reconstructed as two
jets. No electrons or muons with energies above $20 \GeV$ are allowed
in order to reduce the background from W-boson pair production and
subsequent semi-leptonic decay. Events with photons with energy above $20\GeV$
are rejected so as to reduce background from the $\epem\ra\qqbar$
process with a hard initial-state-radiation (ISR) photon. The
background from hadron production in two-photon collisions is
suppressed by limiting the energy deposition in the low-angle
calorimeters and by rejecting events with a two-jet mass, $m_{\rm
qq}$, below $10 \GeV$. A low-mass selection is applied for $m_{\rm
qq}<50 \GeV$ and a high-mass selection otherwise.

The low-mass selection removes background from hadron production in
two-photon collision by requiring at least one jet to point more than
0.3~rad away from the beam axis. The $\epem\ra\qqbar$ process results
in two back-to-back jets in the plane transverse to the beams and, if
no ISR occurred, in two jets which are also back-to-back in space. It
is strongly suppressed by requiring $\rm\Delta\phi<3\,rad$ and
$\rm\Delta\psi<3\,rad$. The requirements $\rm\Delta\phi>1.8\,rad$
and $\rm\Delta\psi>1.8\,rad$ remove background from the $\epem\ra\rm
W e \nu$ process and semi-leptonic decays of W-boson pairs as the
W-boson boost results in opening angles smaller than those for the
signal. Large missing momentum is due to the production of a Z boson
and is tagged by requiring $p_{\rm mis}>0.3~\sqrt{s}$ and $p_{\rm
t}>0.3\sqrt{s}$.

The high-mass selection accepts events with a higher multiplicity and
a missing mass, $m_{\nu\nu}=\sqrt{(\sqrt{s}-E_{\rm vis})^2-p_{\rm
mis}^2}$, compatible with $m_{\rm Z}$, $78\GeV < m_{\nu\nu}
<115\GeV$. Background from W-boson pair production and the
$\epem\ra\qqbar$ process is reduced by requiring $8\GeV<p_{\rm
t}<40\GeV$. The normal to the plane of the two jets must not point
more than 1.5~rad away from the beam axis. In addition, the event
thrust must be greater than $0.78-0.88$ and $\Delta\psi>2-2.5$~rad,
depending on $\sqrt{s}$. The $\epem\ra\qqbar$ process with hard ISR
photons is further reduced by requiring $p_\|<30-42\GeV$.

Figure~\ref{fig:2}a shows the distribution of $\Delta\phi$ for both
the low- and high-mass selections. The residual background is mostly
due to W-boson pair production and  the $\epem\ra\rm W e \nu$
process.

\subsection{The \boldmath{$\qqll$} channel}

The signature of the $\epem\ra\qqll$ process is two hadronic jets and
either an electron or a muon pair. This signature can also arise from
four-fermion events outside the signal definition. Other sources of
background are W-boson pair production and the $\epem\ra\qqbar$
process with leptons coming from heavy-quark decays.

High-multiplicity events with two electrons or two muons are selected
and the remaining calorimetric clusters are reconstructed as two
jets. The measured energies and momenta of the two jets and the two
leptons are varied within their resolutions to best fit the hypotheses
of energy and momentum conservation. This kinematic fit 
improves the resolution of the jet-energy measurements.

Events with W-boson pair production and
semi-leptonic decay have missing momentum due to the neutrinos and are
suppressed by requiring $p_{\rm t}/E_{\rm vis}<0.35$ and $p_\|/E_{\rm
vis}<0.35$. A low-mass selection is applied if $500\GeV^2<m_{\rm
qq}\times m_{\ell\ell}<4000\GeV^2$, where $m_{\rm qq}$ is the two-jet
mass and $m_{\ell\ell}$ the mass of the lepton pair. A high-mass
selection covers the range $m_{\rm qq}\times
m_{\ell\ell}>4000\GeV^2$. Events with $m_{\rm qq}\times
m_{\ell\ell}<500\GeV^2$ exhibit a large background contamination and
are not further considered. The selection criteria depend on the
flavour of the leptons and on $\sqrt{s}$.

\begin{figure}
  \begin{center}
    \begin{tabular}{cc}
      \mbox{\includegraphics*[width=0.45\textwidth]{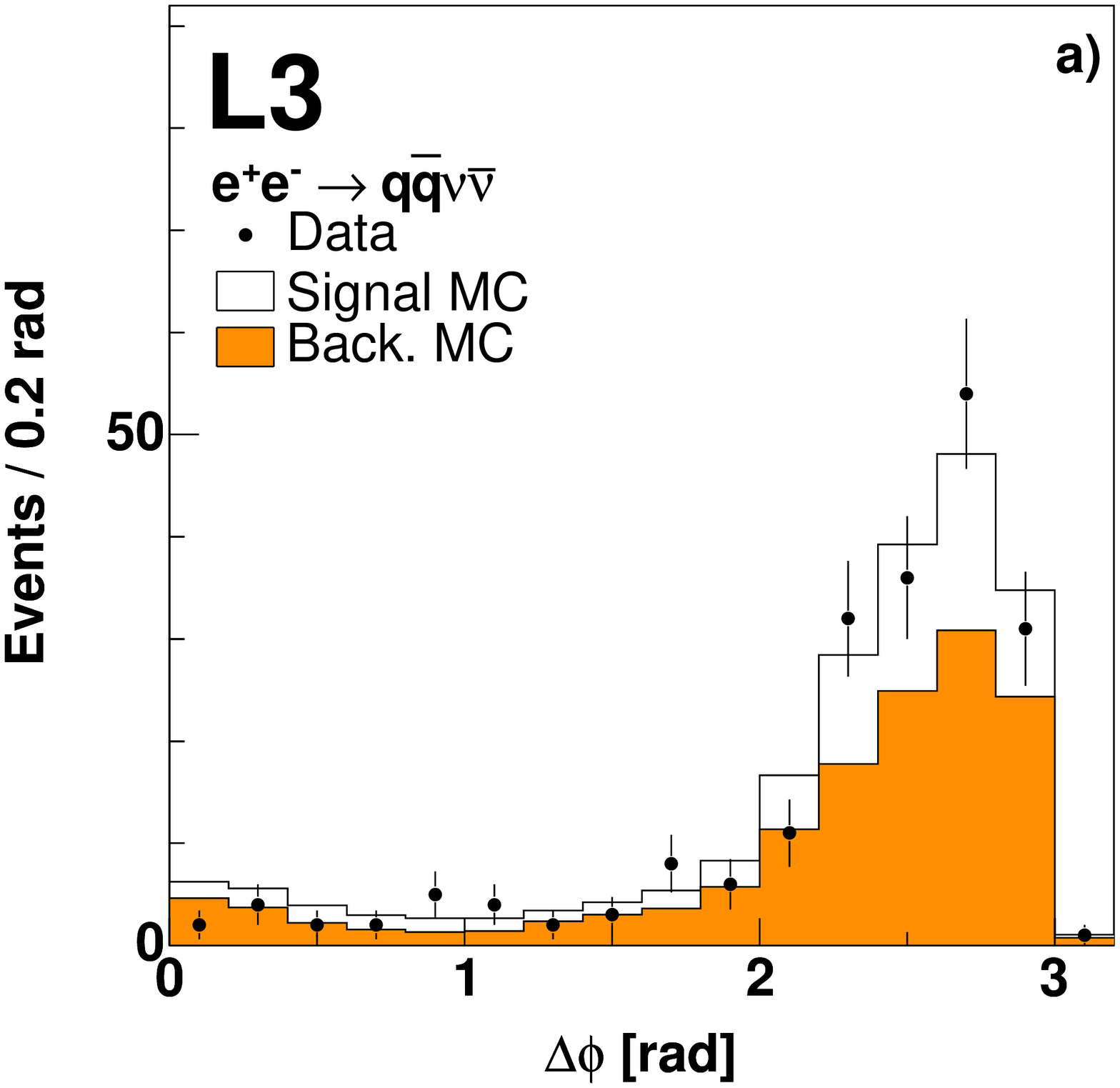}} & 
      \mbox{\includegraphics*[width=0.45\textwidth]{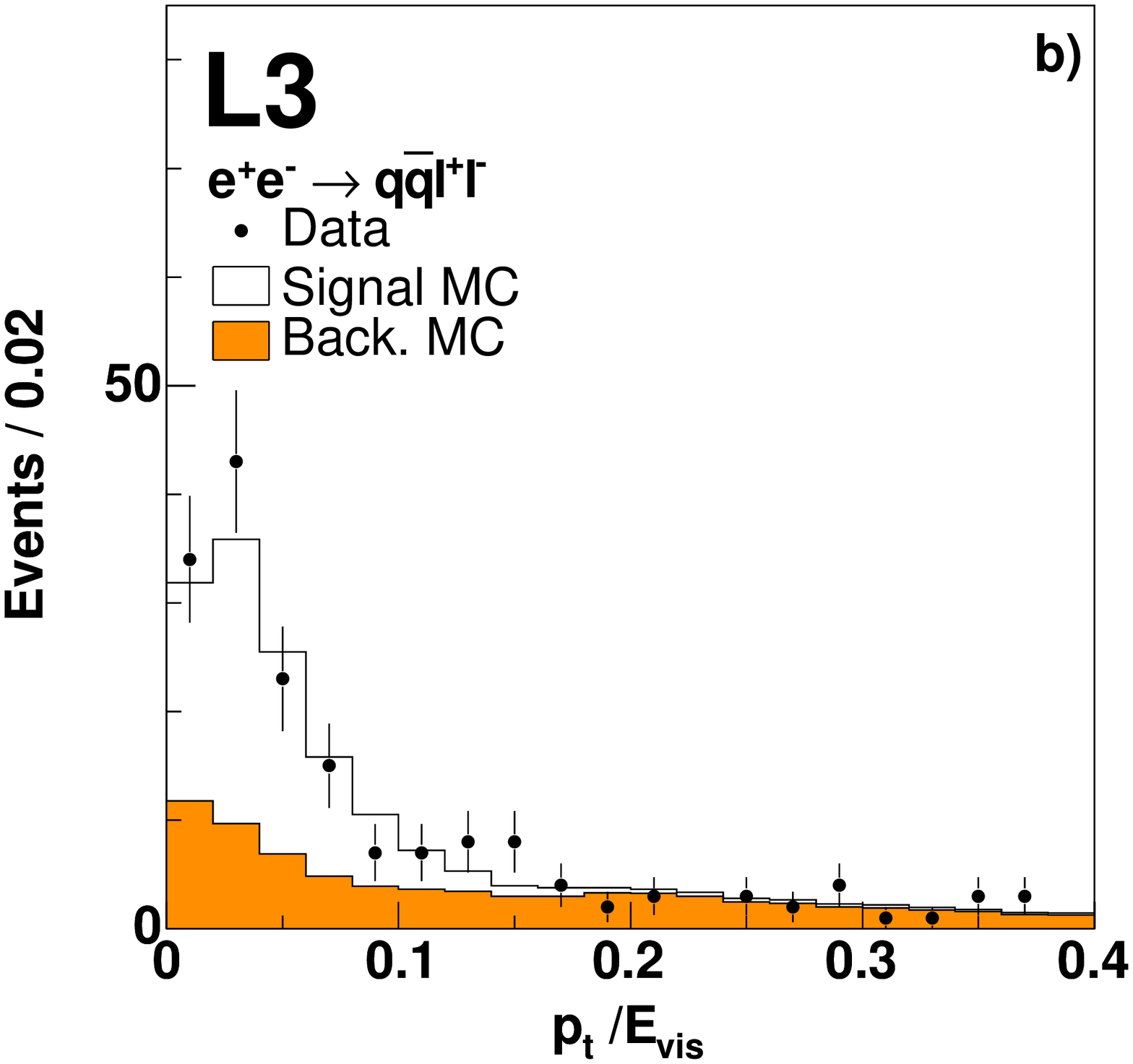}} \\
      \mbox{\includegraphics*[width=0.45\textwidth]{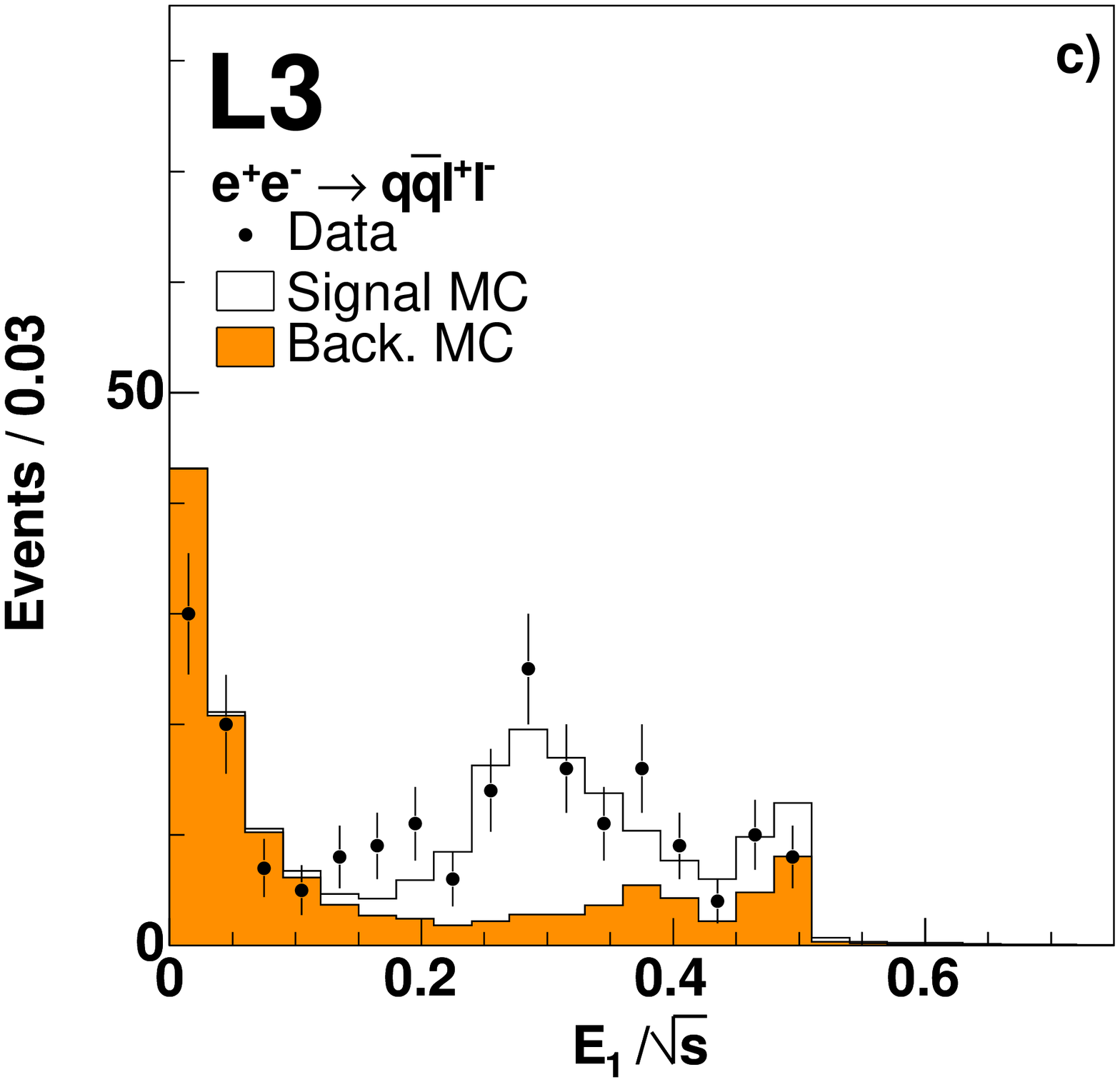}} & 
      \mbox{\includegraphics*[width=0.45\textwidth]{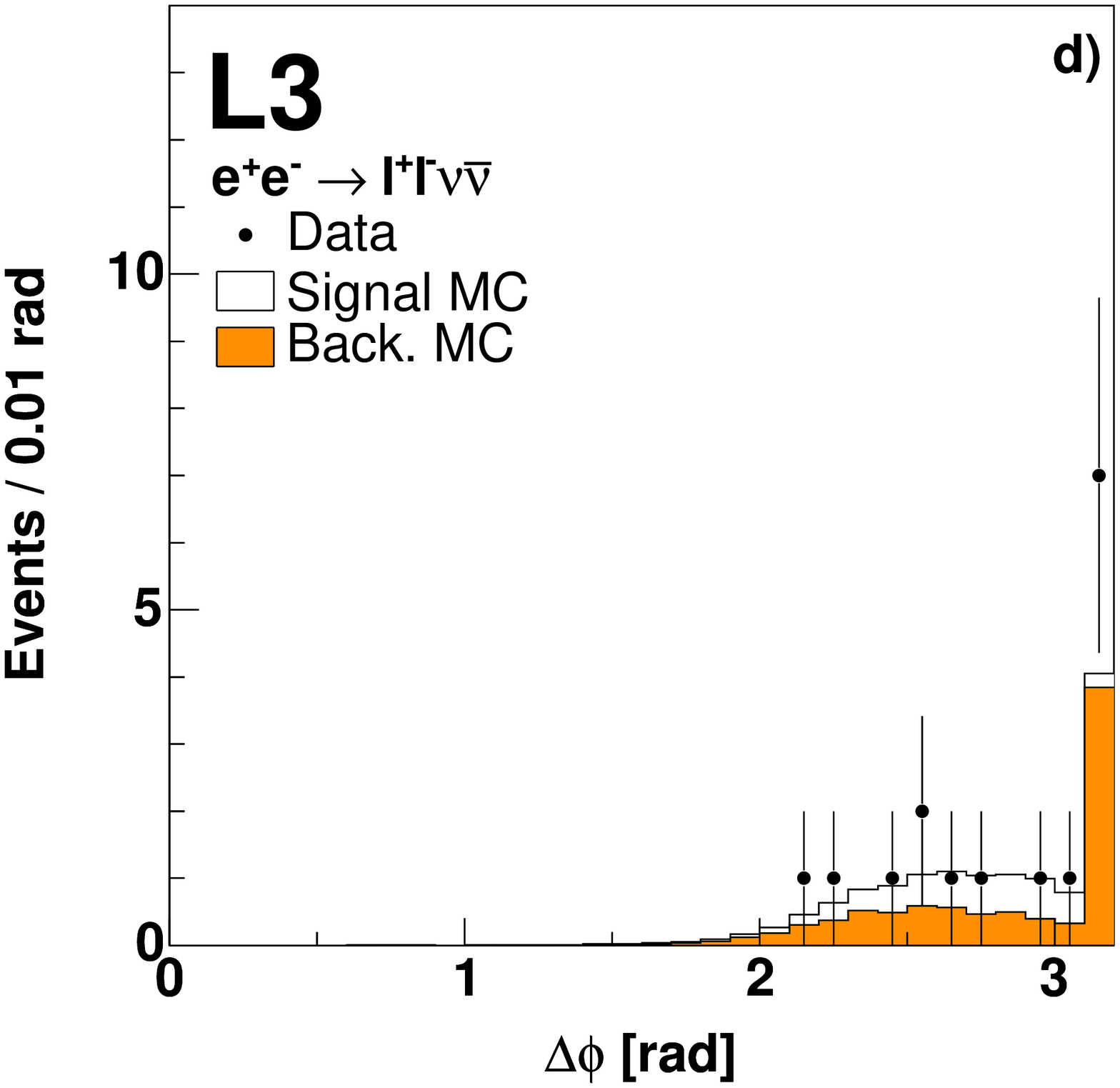}} \\
    \end{tabular}
    \caption{Distribution of selection variables for the different
      channels.  a) The angle in the plane transverse to the beams for the
      two jets of the $\qqnn$ channel, b) the transverse momentum
      normalised to the visible energy for the $\qqll$ channel, c) the
      energy of the most energetic lepton normalised to $\sqrt{s}$ for the
      $\qqll$ channel and d) the angle in the plane transverse to the beams
      for the two leptons of the $\llnn$ channel. In each plot, all selection
      cuts are applied with the exception of that on the shown variable.}
    \label{fig:2}
  \end{center}
\end{figure}

In order to reduce the background from W-boson pair
production, the low-mass analysis requires $p_{\rm t}/E_{\rm
vis}<0.12-0.14$ for electrons and $p_{\rm t}/E_{\rm vis}<0.22-0.30$
for muons. The energy of the most energetic lepton, $E_1$, is required
to satisfy $E_1/\sqrt{s}>0.12-0.15$ in order to remove leptons from
heavy-quark decays. For electrons, the energy of the least energetic
lepton, $E_2$, must satisfy $E_2/\sqrt{s}>0.07-0.15$.  The high-mass analysis
requires $E_2/\sqrt{s}>0.06-0.10$ and $p_{\rm t}/E_{\rm
vis}<0.11-0.14$ for electrons and $E_1/\sqrt{s}>0.10-0.16$ and $E_{\rm
vis}/\sqrt{s}>0.5-0.7$ for muons.  Figures~\ref{fig:2}b
and~\ref{fig:2}c, show the distributions of $p_{\rm t}/E_{\rm vis}$
and $E_1/\sqrt{s}$.

The residual background in both channels is due to events from
four-fermion production outside the signal definition, from W-boson
pair production and from the $\epem\ra\qqbar$
process. 

\subsection{The \boldmath{$\llnn$} channel}

The signature of the $\epem\ra\llnn$ process is an electron or muon
pair and large missing energy, mostly due to a Z boson decaying into
neutrinos. The most important backgrounds are lepton pair production
with a hard ISR photon and W-boson pair production.

Events with just one or two tracks associated to two identified
electrons or muons are selected. To reduce background from
annihilation and two-photon lepton pair production, no large
energy deposition is allowed in the low polar-angle
calorimeters. Events with $m_{\ell\ell}<10\GeV$ are removed from the
sample. The remaining events are considered by three overlapping
selections, according to the value of $m_{\ell\ell}$. Background from
lepton pair production results in low-$p_{\rm t}$ events with
back-to-back leptons and is reduced by requiring $p_{\rm
t}/\sqrt{s}>0.1-0.3$ and $\Delta\phi<2.6-3.1$~rad as shown in
Figure~\ref{fig:2}d. The background is strongly suppressed by
requiring the recoil mass to be compatible with $m_{\rm Z}$. Finally,
the signal purity is enhanced by requiring $E_{\rm
vis}/\sqrt{s}>0.4-0.5$ and $E_1/\sqrt{s}>0.2-0.4$.

The remaining background is almost entirely due to $\epem\ra\llnn$
events outside the signal definition criteria.

\subsection{The \boldmath{$\llll$} channel}

The signature of the $\epem\ra\llll$ process is four leptons of which
at least one pair originates from a Z boson or a low-mass virtual
photon. These configurations are found in background from four-fermion
events outside the signal definition criteria and lepton pair
production with additional radiative photons, which mimic electrons in
the detector.  

Events with three or four tracks and four identified leptons are
selected if $E_{\rm vis}/\sqrt{s}>0.4$.  All same-flavour lepton pairs
are considered and their mass is calculated. If at least one pair has
a mass above $40\GeV$, the event is considered by a high-mass
selection which starts from the lepton pair with mass closer to
$m_{\rm Z}$. A low-mass selection, aimed to identify events with a
lepton pair originating from a low-mass off-mass-shell photon, is
applied if at least one lepton pair has a mass below $60\GeV$. The
low-mass selection starts from the lepton pair with the lowest
mass. An event can be considered and selected by both selections.

The high-mass selection requires the lepton pair to have $\rm
1.55~rad<\Delta\psi<3.10~rad$, where the upper cut removes
non-radiative fermion-pair events. The other two leptons must satisfy
$\rm 0.1~rad<\Delta\phi<2.3$~rad.  Radiative lepton-pair production is
further suppressed by requiring $p_\|/\sqrt{s}<0.4$. Four-fermion
background is reduced by requiring the masses of both pairs to be
below $120\GeV$.

The low-mass selection requires the lepton pair to have $\rm 0.1~rad
<\Delta\psi<2.8~rad$, while the two other leptons must satisfy
$\rm\Delta\phi<1.0~rad$ and $\rm 1.45~rad<\Delta\psi<3.05~rad$.
Radiative lepton-pair production is suppressed by requiring
$p_\|/\sqrt{s}<0.36$, $p_{\rm t}/\sqrt{s}>0.06$ and by upper cuts on
the lepton energies. Four-fermion events outside the signal definition
are removed by requiring the mass of the second pair to be in the
range $14-152\GeV$.

The residual background originates in equal parts from lepton pair
production and four-fermion events outside the signal definition
criteria. Some contribution from lepton production in two-photon
collisions is observed for final states with electrons.

%
\section{Results}
%

\subsection{The \boldmath{$\epem\ra f\bar{f}f'\bar{f}'$} process}

\begin{figure}
  \begin{center}
      \mbox{\includegraphics*[width=0.9\textwidth]{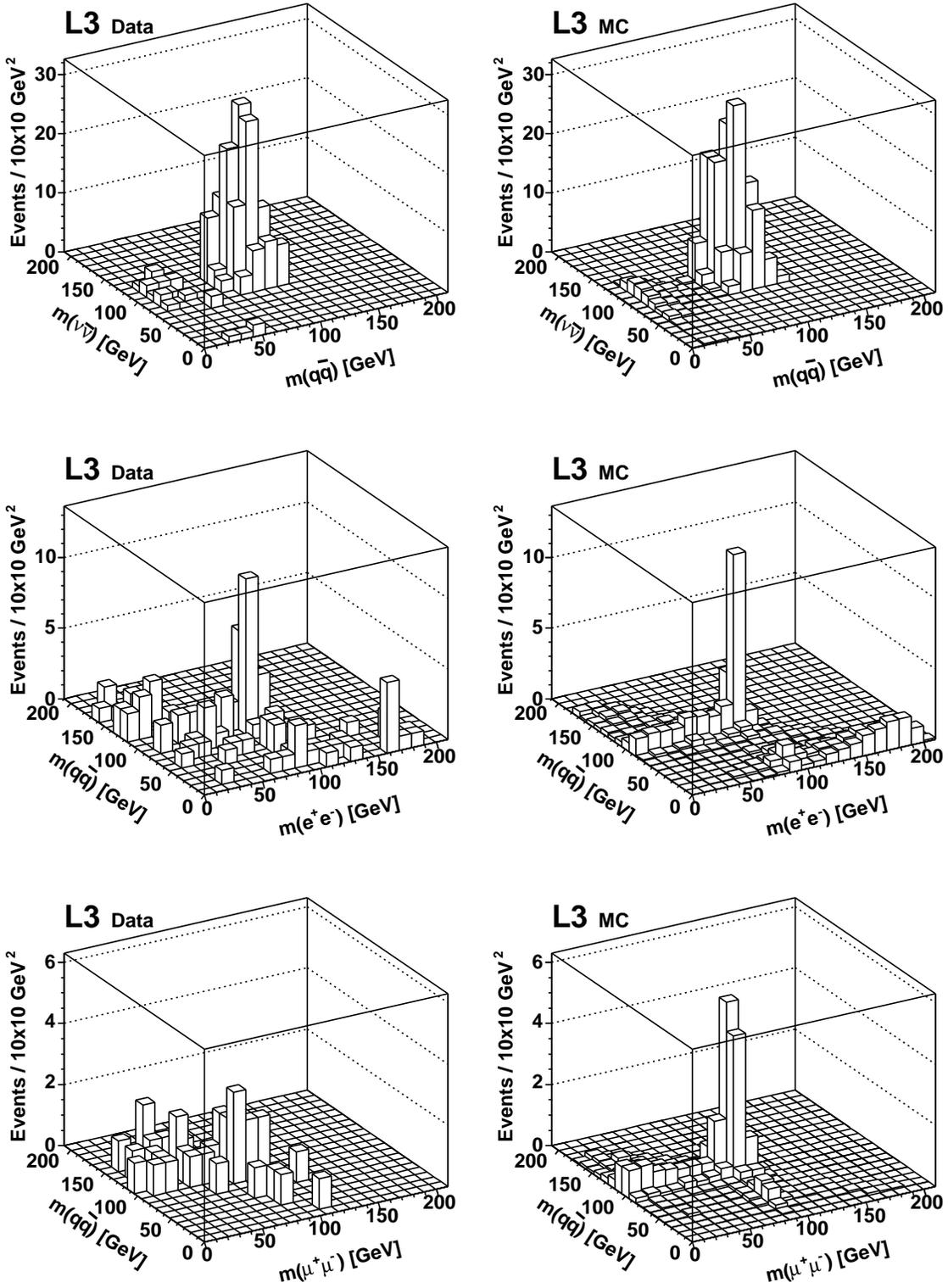}}
    \caption{Distributions of the missing mass {\it vs.} the
      hadron mass for the $\qqnn$ channel, first row, and the lepton
      mass {\it vs.} the hadron mass for the $\qqee$ channel, second
      row, and $\qqmm$ channel, third row. The left-hand side plots
      represent the data, the right-hand side ones the Monte Carlo
      predictions.}
    \label{fig:3}
  \end{center}
\end{figure}

Figure~\ref{fig:3} presents two-dimensional plots of the $m_{\nu\nu}$
and $m_{\rm qq}$ masses for the $\qqnn$ final state and the
$m_{\ell\ell}$ and $m_{\rm qq}$ masses for the $\qqll$ final states.
The data exhibit  contributions from Z-boson pair production,
final states with a Z boson and an off-mass-shell virtual photon and
the continuum. A good agreement is observed with the Monte Carlo
predictions, for the $\qqnn$ and $\qqee$ final states, with some
fluctuations for the $\qqmm$ final state.

\begin{figure}
  \begin{center}
    \begin{tabular}{cc}
      \mbox{\includegraphics*[width=0.45\textwidth]{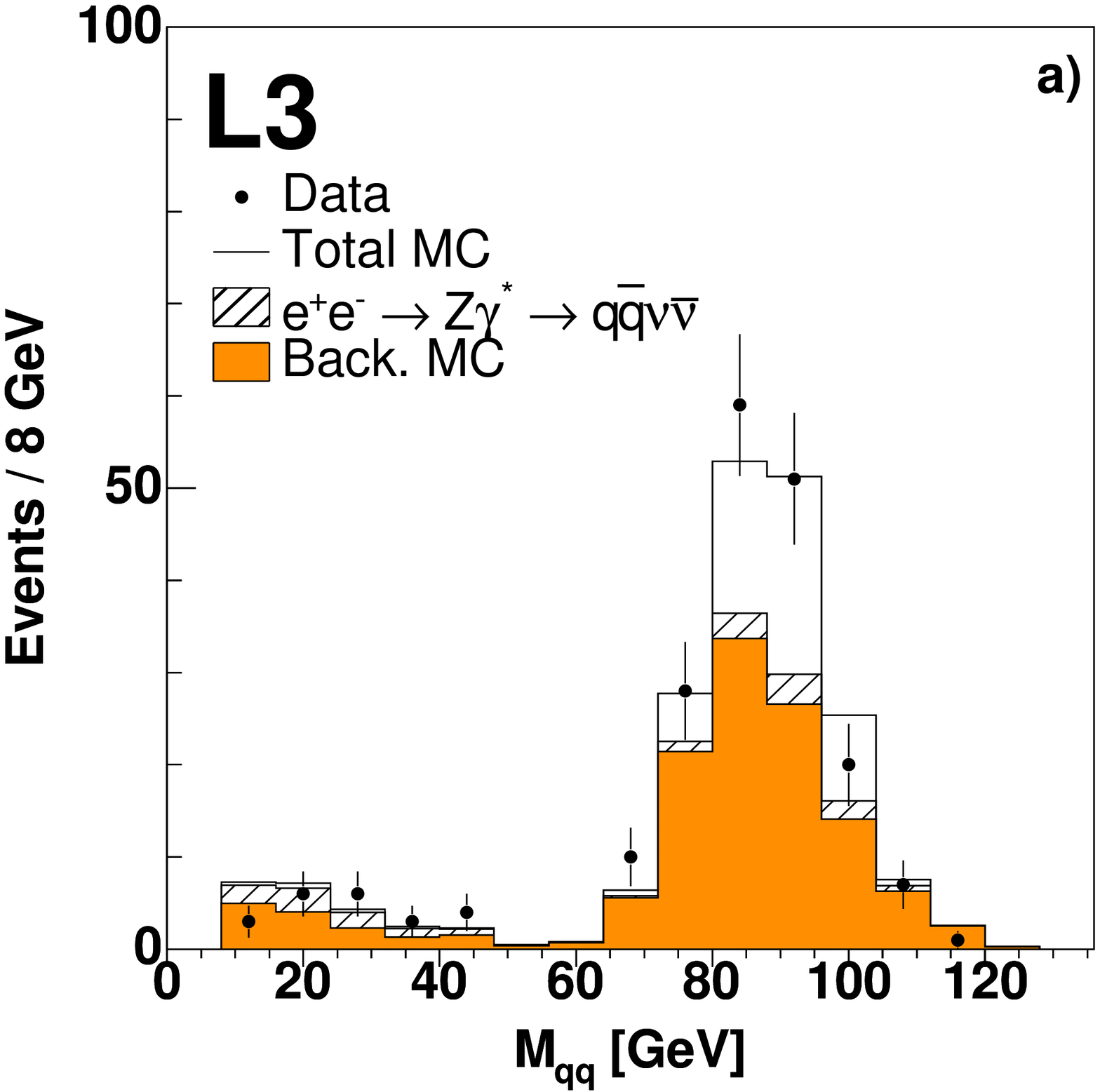}} & 
      \mbox{\includegraphics*[width=0.45\textwidth]{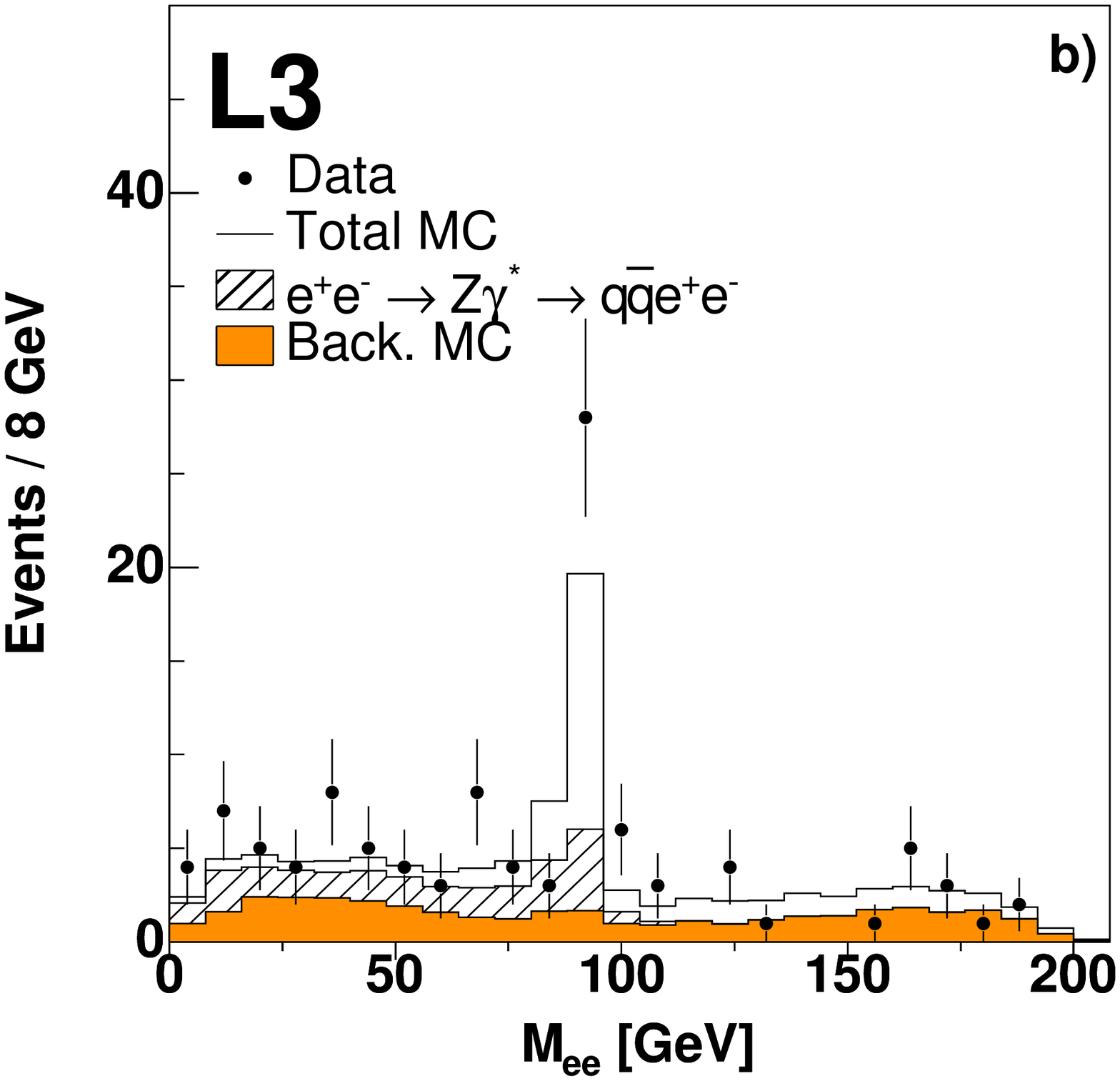}} \\
      \mbox{\includegraphics*[width=0.45\textwidth]{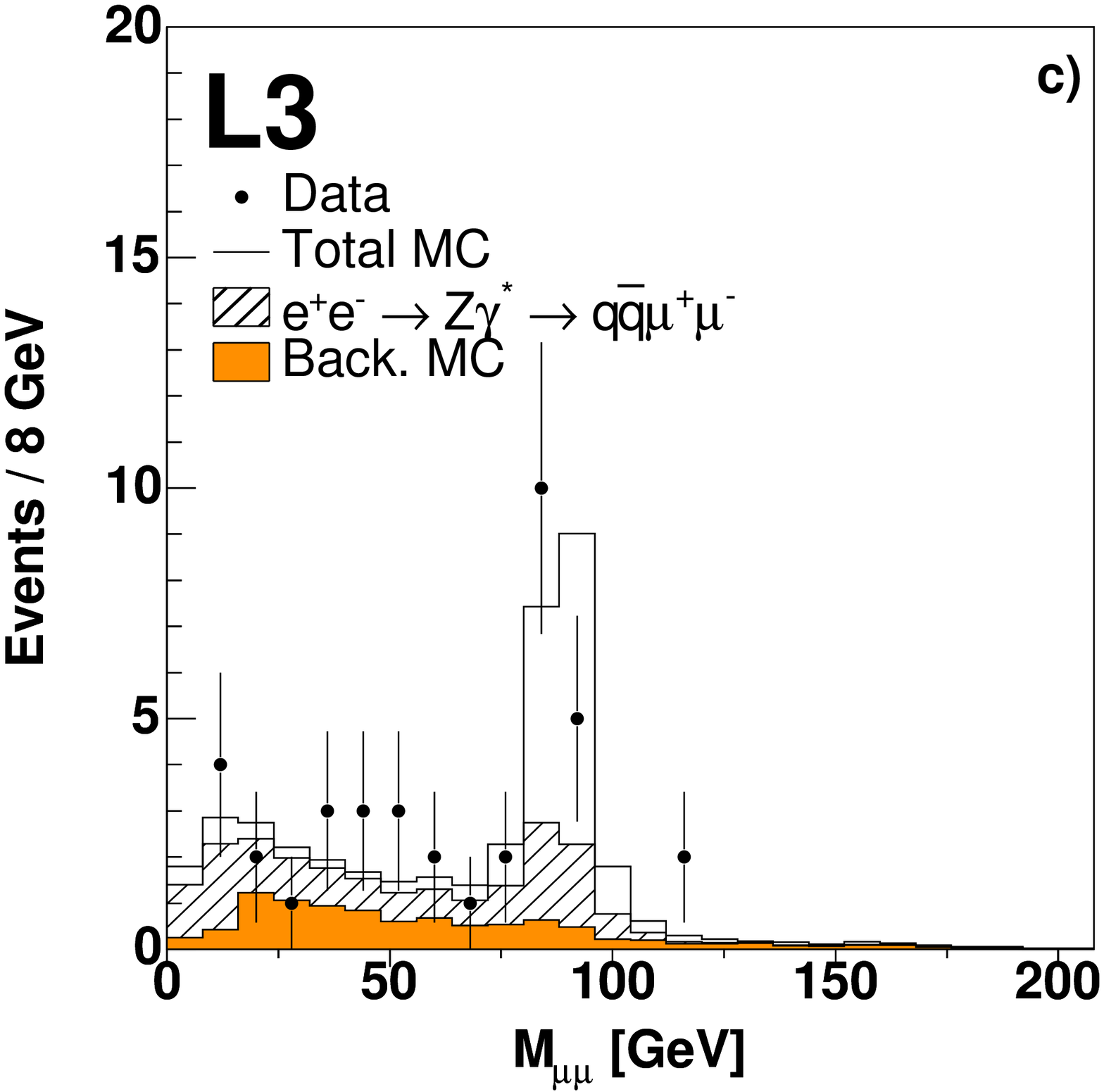}} & 
      \mbox{\includegraphics*[width=0.45\textwidth]{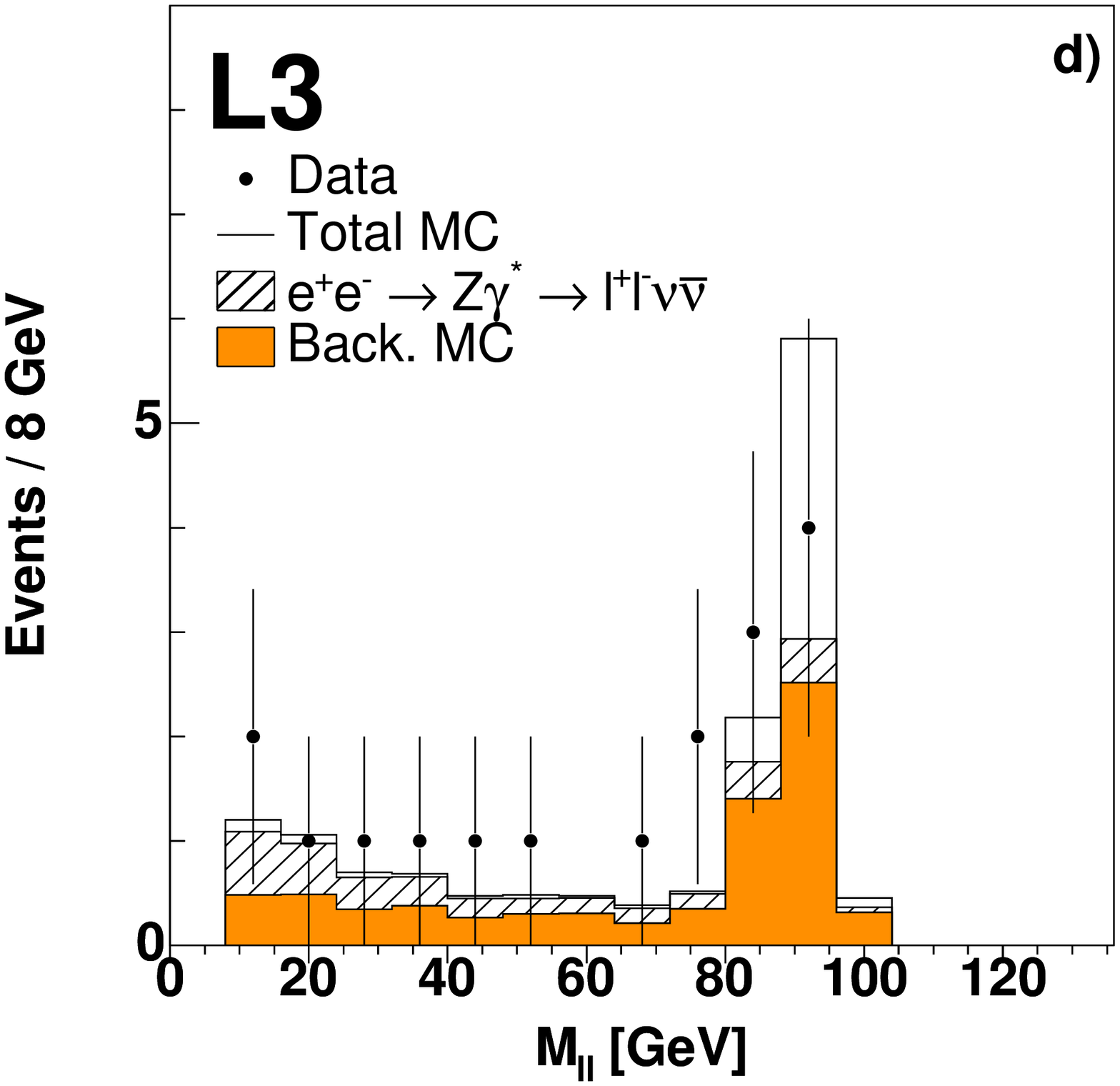}} \\
      \mbox{\includegraphics*[width=0.45\textwidth]{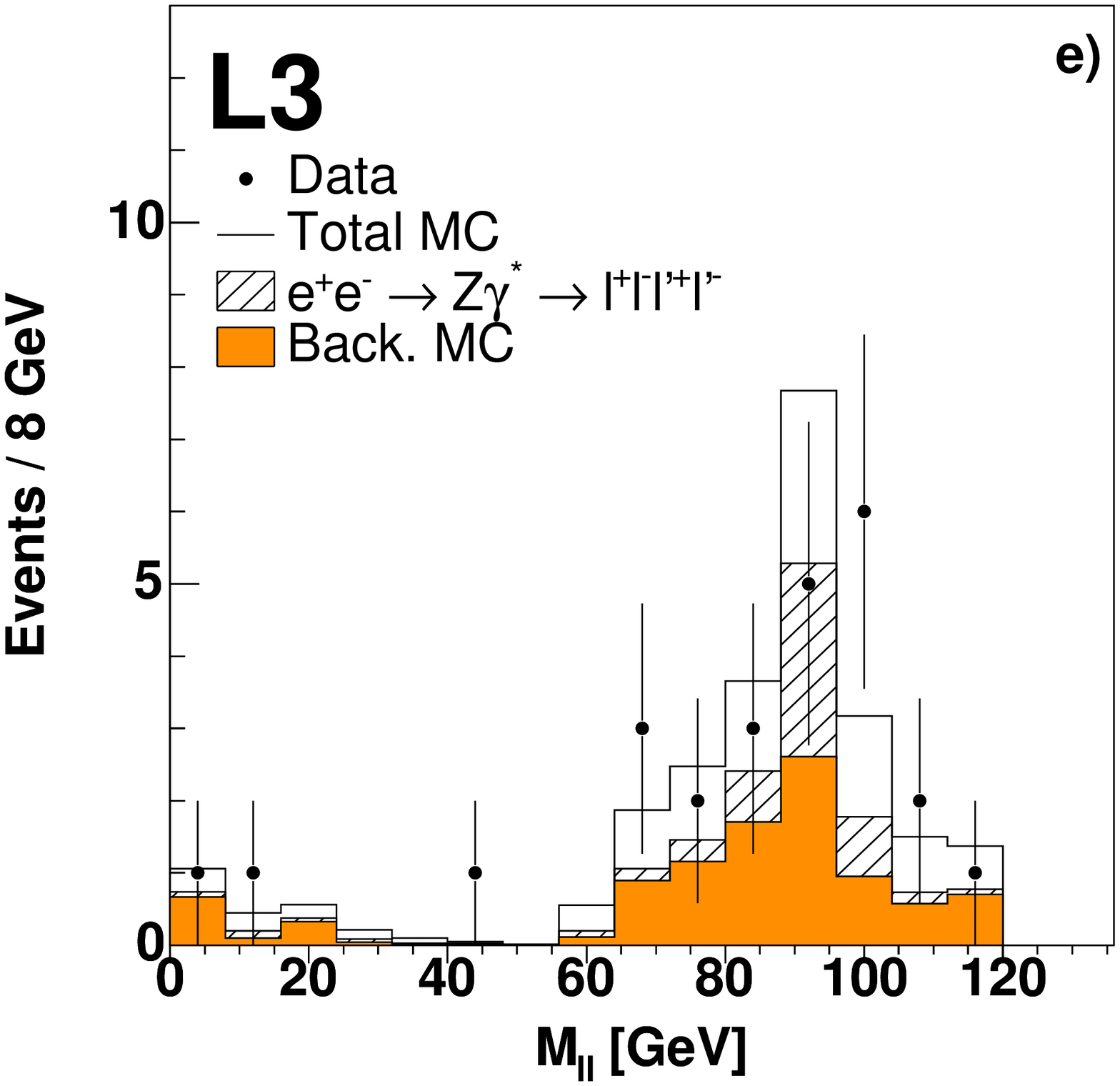}} & \\
    \end{tabular}			
    \caption{ 
      Distributions of a) the mass of the hadron system in the $\qqnn$
      channel; the mass of the lepton pair in  the b) $\qqee$  c)
      $\qqmm$  and d)  $\llnn$ channels and  e) the mass of the
      selected lepton pair in the $\llll$ channel.}
    \label{fig:4}
  \end{center}
\end{figure}

The cross sections for neutral-current four-fermion production are derived
from the $m_{\ell\ell}$ spectra for all channels with leptons and the
$m_{\rm qq}$ spectrum for the $\qqnn$ final state, shown in
Figure~\ref{fig:4}.  The background level and shape is fixed to the
Monte Carlo predictions, and the $\epem\ra f\bar{f}f'\bar{f}'$
normalisation is then derived from  the data.  Table~\ref{tab:2}
compares the measured and expected cross sections, determined as the
luminosity-weighted average of the cross sections at each of the
average centre-of-mass energies listed in Table~\ref{tab:1}. A good
agreement is observed.

Several possible sources of systematic uncertainties are considered as
summarised in Table~\ref{tab:3}.  The jet and lepton identification
and reconstruction are affected by the energy scale of the
calorimeters and the accuracy of the track measurements. The analysis
is repeated by varying  the energy scale by $\pm2\%$ and modifying
the lepton selection criteria. The differences are considered as
systematic uncertainties.

Monte Carlo modelling of the detector response is in general
accurate. Possible systematic uncertainties could arise from distortions
in the modelling of tails of distributions used in the event
selection. These are addressed by varying the selection criteria on
the variables susceptible to remove the largest part of the background
by an amount compatible with their resolution. The measurements of the
cross sections are repeated and largest variations are assigned as
systematic uncertainties.

Small effects from the limited signal and background Monte Carlo
statistics are also considered as systematic uncertainties.

Finally, the uncertainties of the background normalisations are
propagated to the measured cross sections. A variation of $\pm10\%$ is
assumed for neutral-current four-fermion events generated with
EXCALIBUR which fail the signal identification criteria and for the
$\epem\ra\rm W e \nu$ process, $\pm2\%$ on fermion pair production,
$\pm0.5\%$ on W-boson pair production, and $\pm25\%$ and $\pm50\%$ on
lepton and hadron production in
two-photon collisions, respectively.

\subsection{The \boldmath{$\epem \ra{\rm Z} \gamma^* \ra f\bar{f}f'\bar{f}'$} process}

The cross sections of the $\epem \ra{\rm Z} \gamma^* \ra
f\bar{f}f'\bar{f}'$ process are determined with the same procedure as
described above.  Signal Monte Carlo events are subjected to the
additional signal definition criteria for the $\epem \ra{\rm Z}
\gamma^* \ra f\bar{f}f'\bar{f}'$ process described in
Table~\ref{tab:0}. The selected events are treated as the $\epem
\ra{\rm Z} \gamma^* \ra f\bar{f}f'\bar{f}'$ signal, illustrated in
Figure~\ref{fig:4}. Events which fail these criteria are considered as
an additional background and their cross section is fixed to the
predictions. The cross sections measured for each channel are
presented in Table~\ref{tab:4}. Systematic uncertainties are assessed
as for neutral-current four-fermion production and listed in
Table~\ref{tab:3}.

By combining the five different channels, the total $\epem \ra{\rm Z}
\gamma^* \ra f\bar{f}f'\bar{f}'$ cross section at $\langle \sqrt{s}
\rangle=196.6\GeV$ is determined to be $0.29\pm 0.05 \pm 0.03~\rm pb$,
where the first uncertainty is statistical and the second systematic,
to be compared with the GRC4F prediction of 0.22~pb.

\subsection{Production of neutral heavy particles}

\begin{figure}
  \begin{center}
    \begin{tabular}{cc}
      \mbox{\includegraphics*[width=0.45\textwidth]{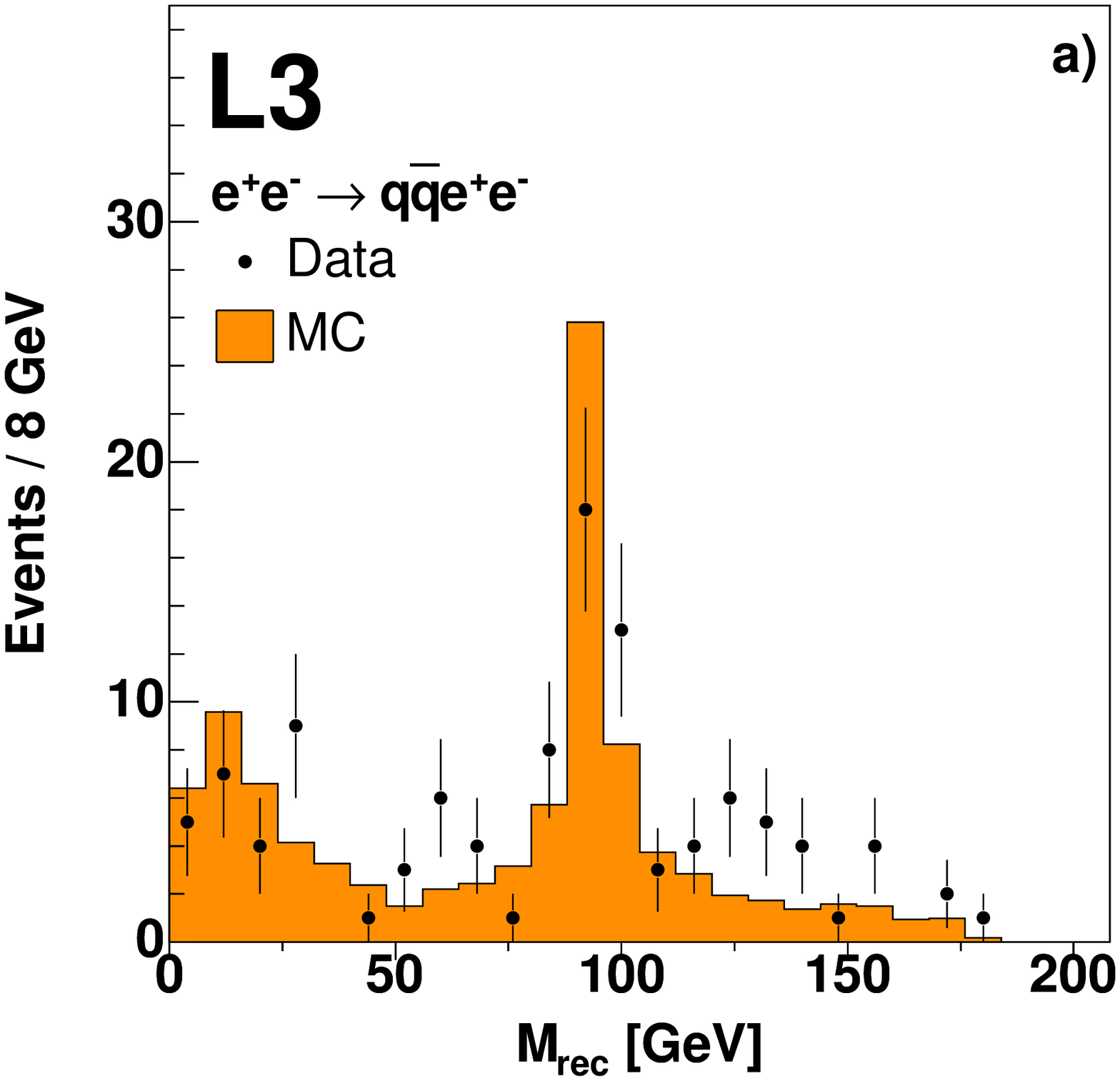}} & 
      \mbox{\includegraphics*[width=0.45\textwidth]{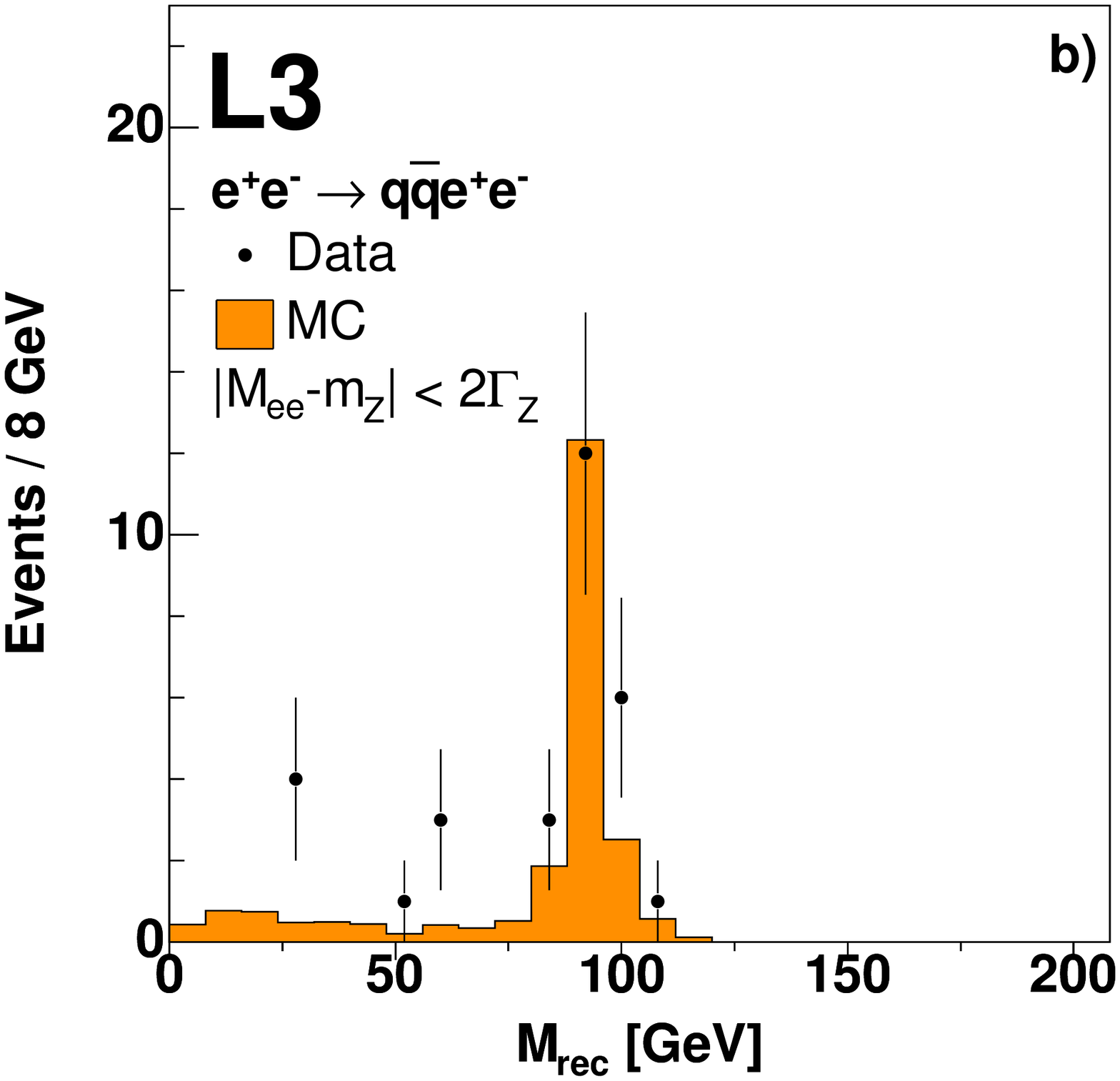}} \\
      \mbox{\includegraphics*[width=0.45\textwidth]{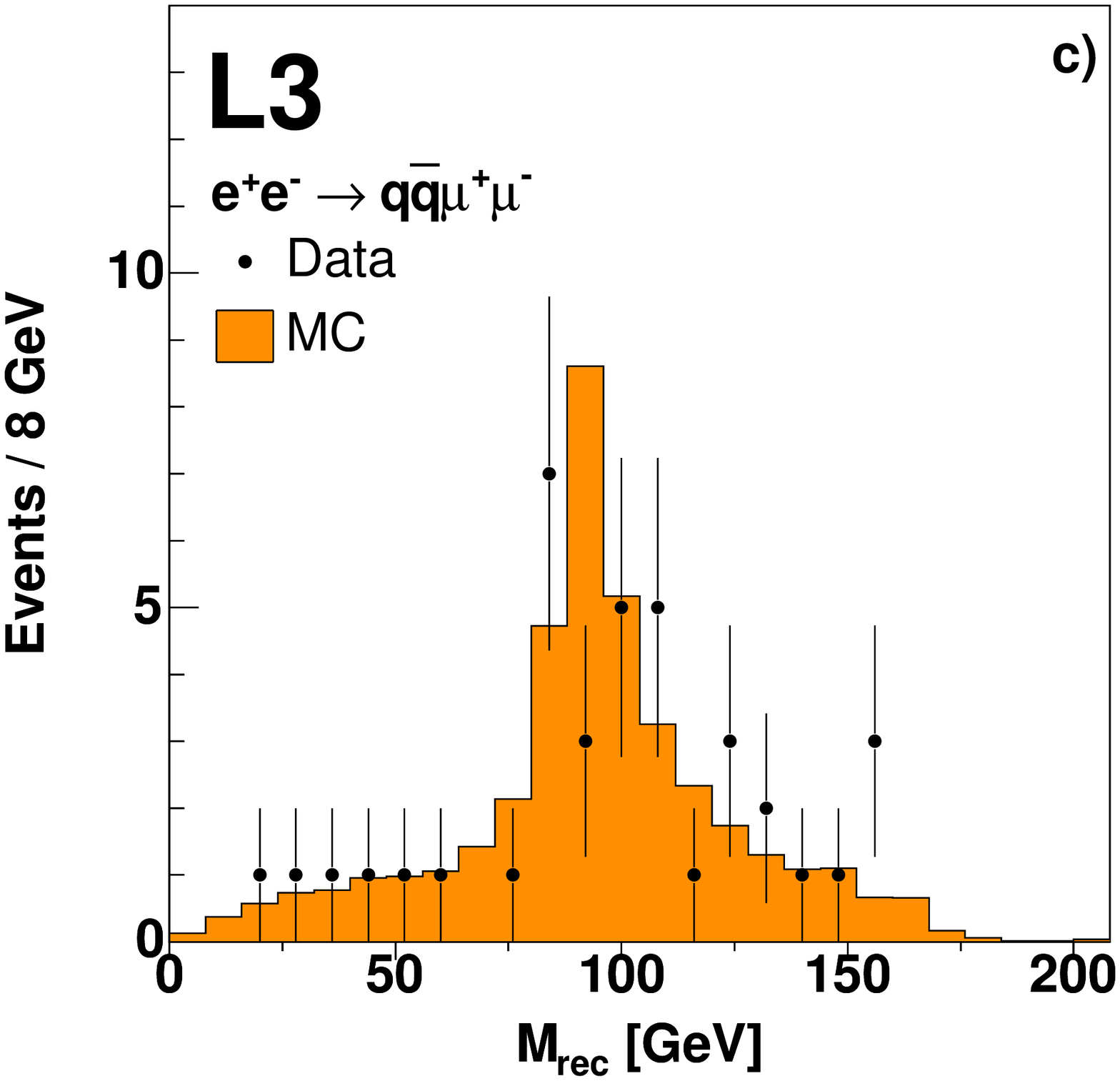}} & 
      \mbox{\includegraphics*[width=0.45\textwidth]{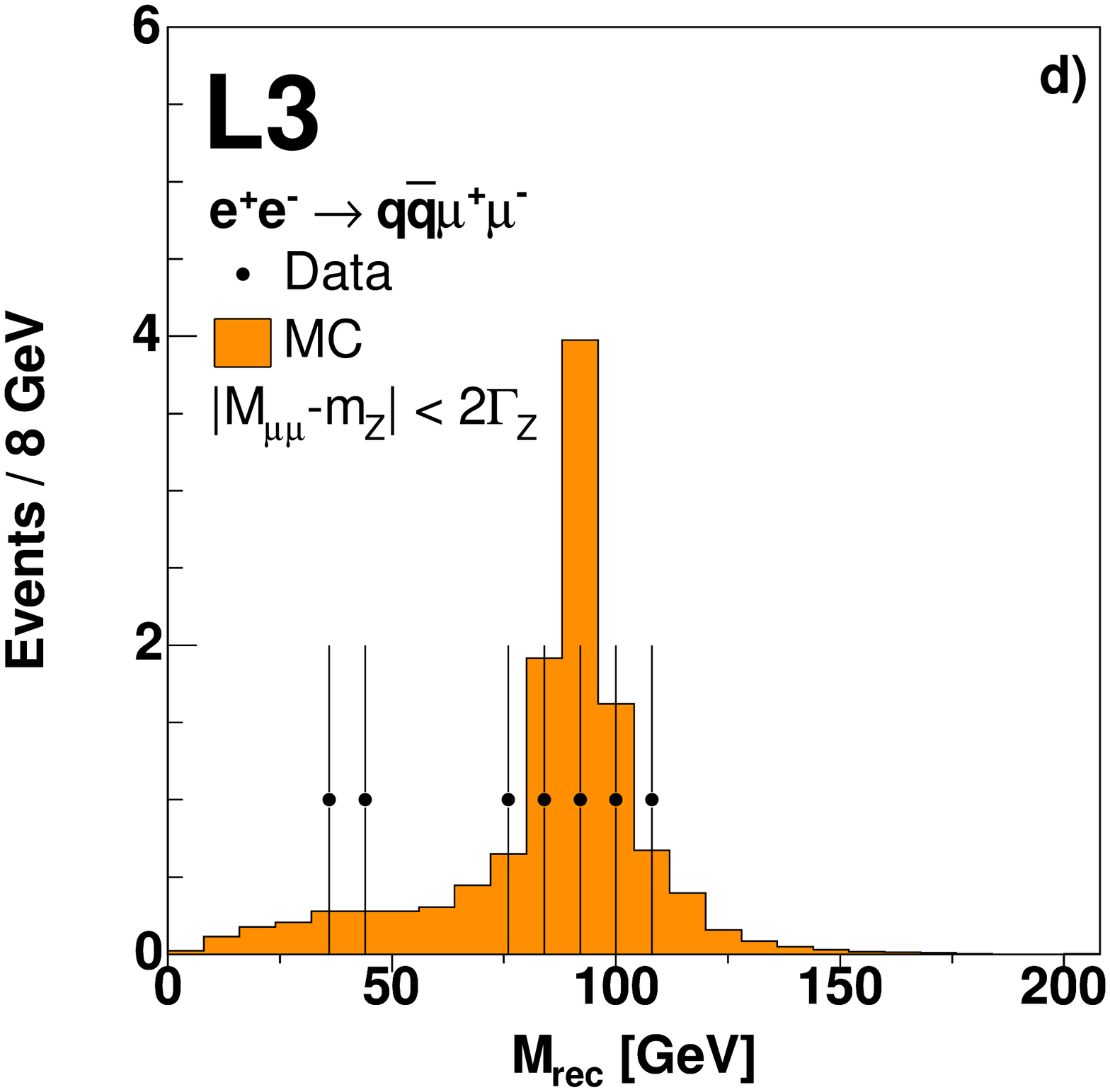}} \\ 
      \mbox{\includegraphics*[width=0.45\textwidth]{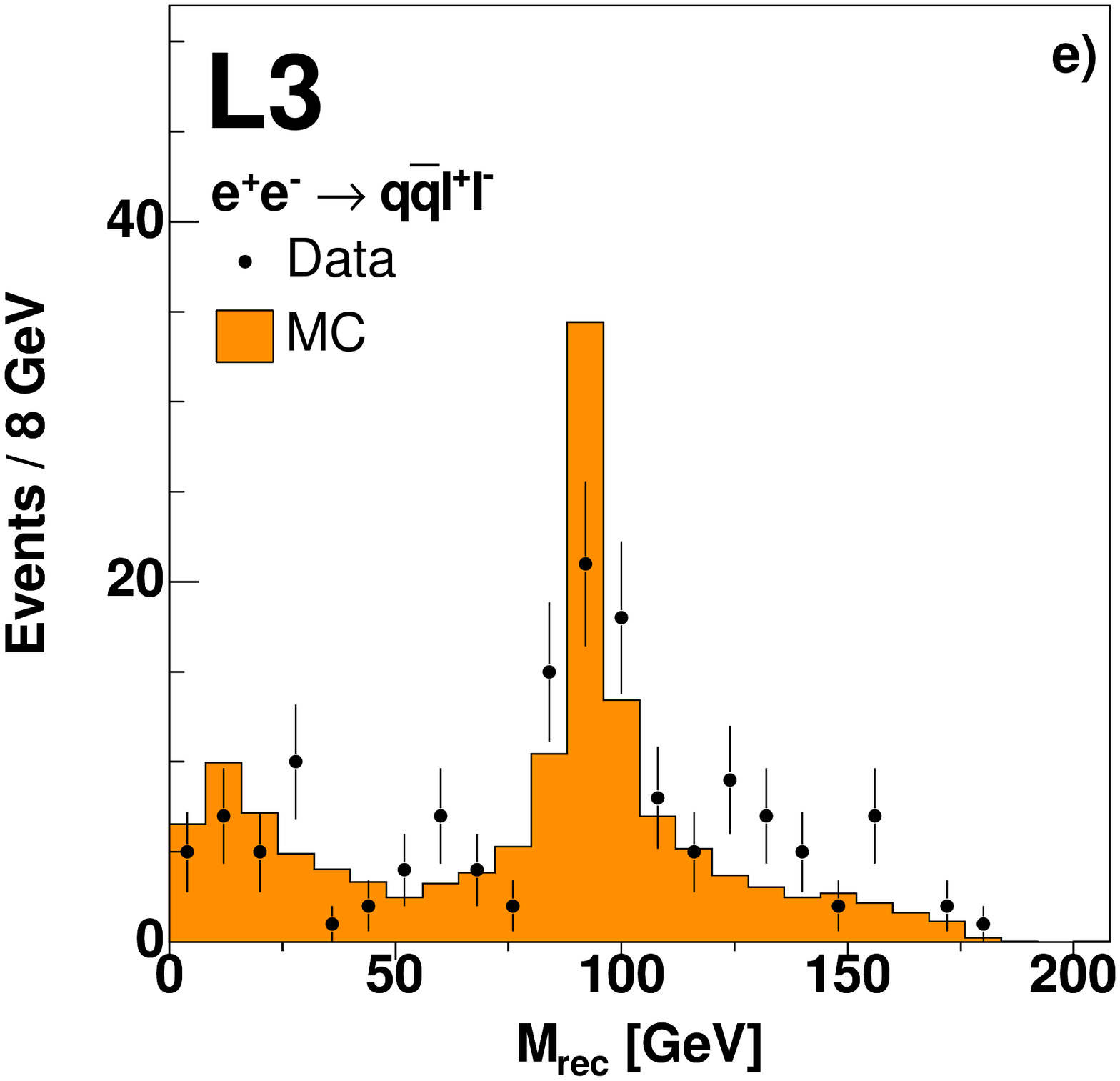}} &
      \mbox{\includegraphics*[width=0.45\textwidth]{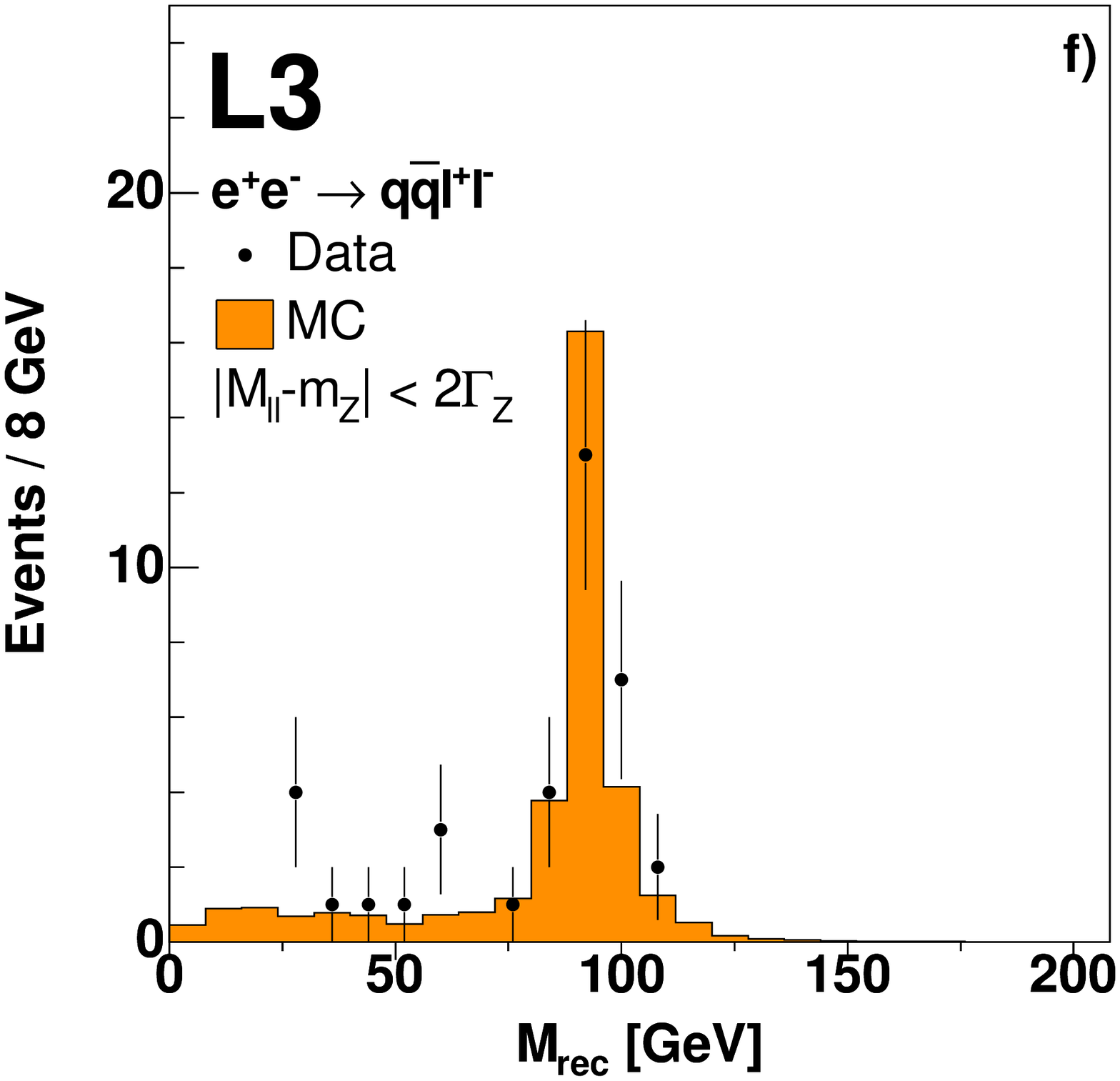}} \\ 
    \end{tabular}
    \caption{Distribution of the recoil mass to the lepton system for
      the a) $\qqee$ and c) $\qqmm$ channels with e) their
      sum. Figures b), d) and f) show the same variables as a), c) and
      e), respectively, if a cut  $|m_{\ell\ell}-m_{\rm Z}|<2\Gamma_{\rm Z}$ is
      applied on the lepton mass.}
    \label{fig:5}
  \end{center}
\end{figure}

Figures~\ref{fig:5}a and~\ref{fig:5}c present the spectra of the mass
recoiling against the electron or muon pairs, respectively, while
Figures~\ref{fig:5}e presents their sum. A new, neutral, heavy
particle decaying into hadrons would manifest as a peak in these
distributions. All distributions agree with the Monte Carlo
predictions: apart from the Z-boson peak, no other significant
structure is observed. The study is narrowed to the case in which the
mass of the lepton pair is compatible with $|m_{\ell\ell}-m_{\rm
Z}|<2\Gamma_{\rm Z}$. Again, no significant deviations from the
Standard Model predictions are observed, as presented in
Figures~\ref{fig:5}b,~\ref{fig:5}d and~\ref{fig:5}f.

\section*{Summary}

\begin{figure}
  \begin{center}
    \begin{tabular}{c}
    \includegraphics*[width=0.6\textwidth]{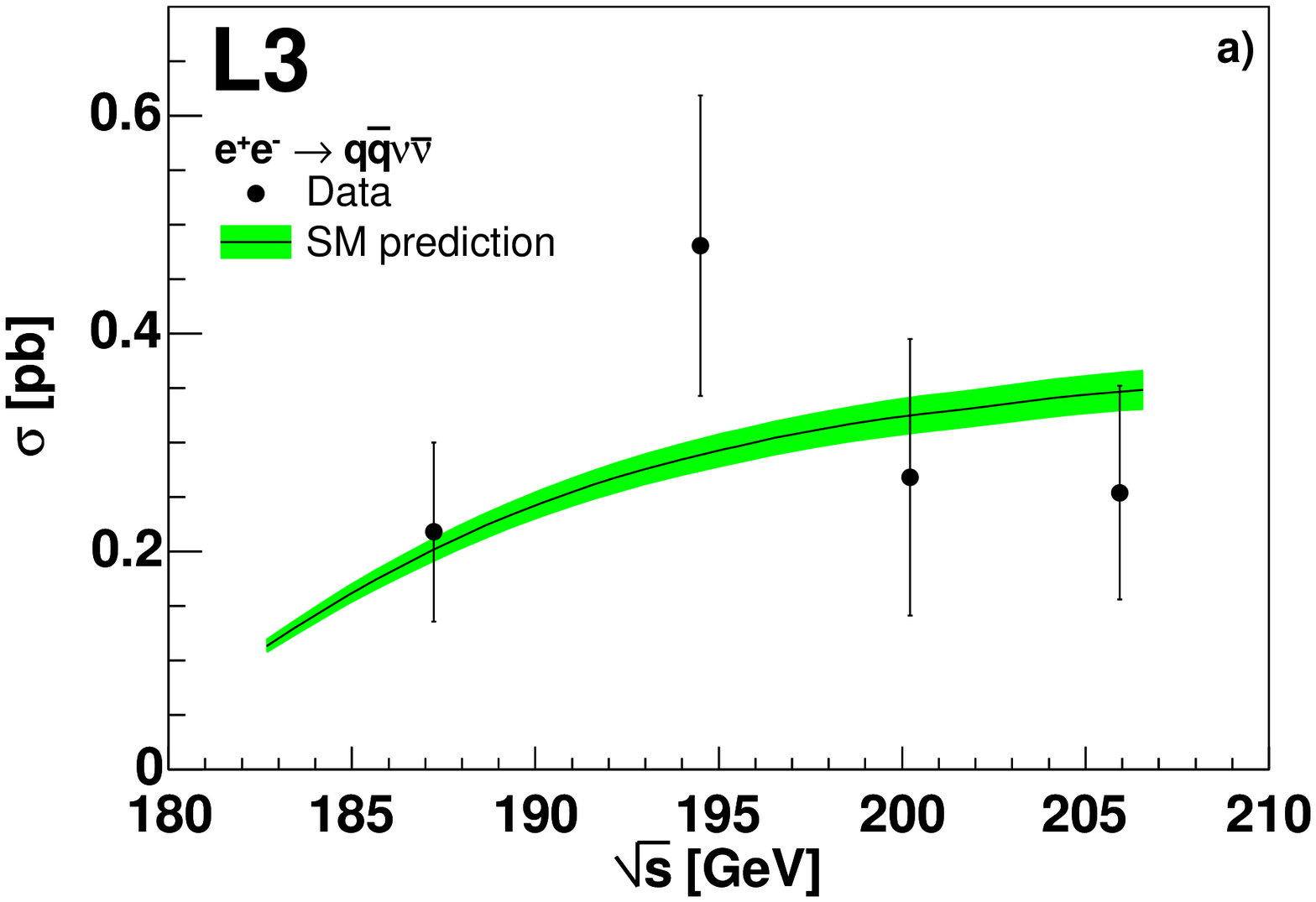}\\
    \includegraphics*[width=0.6\textwidth]{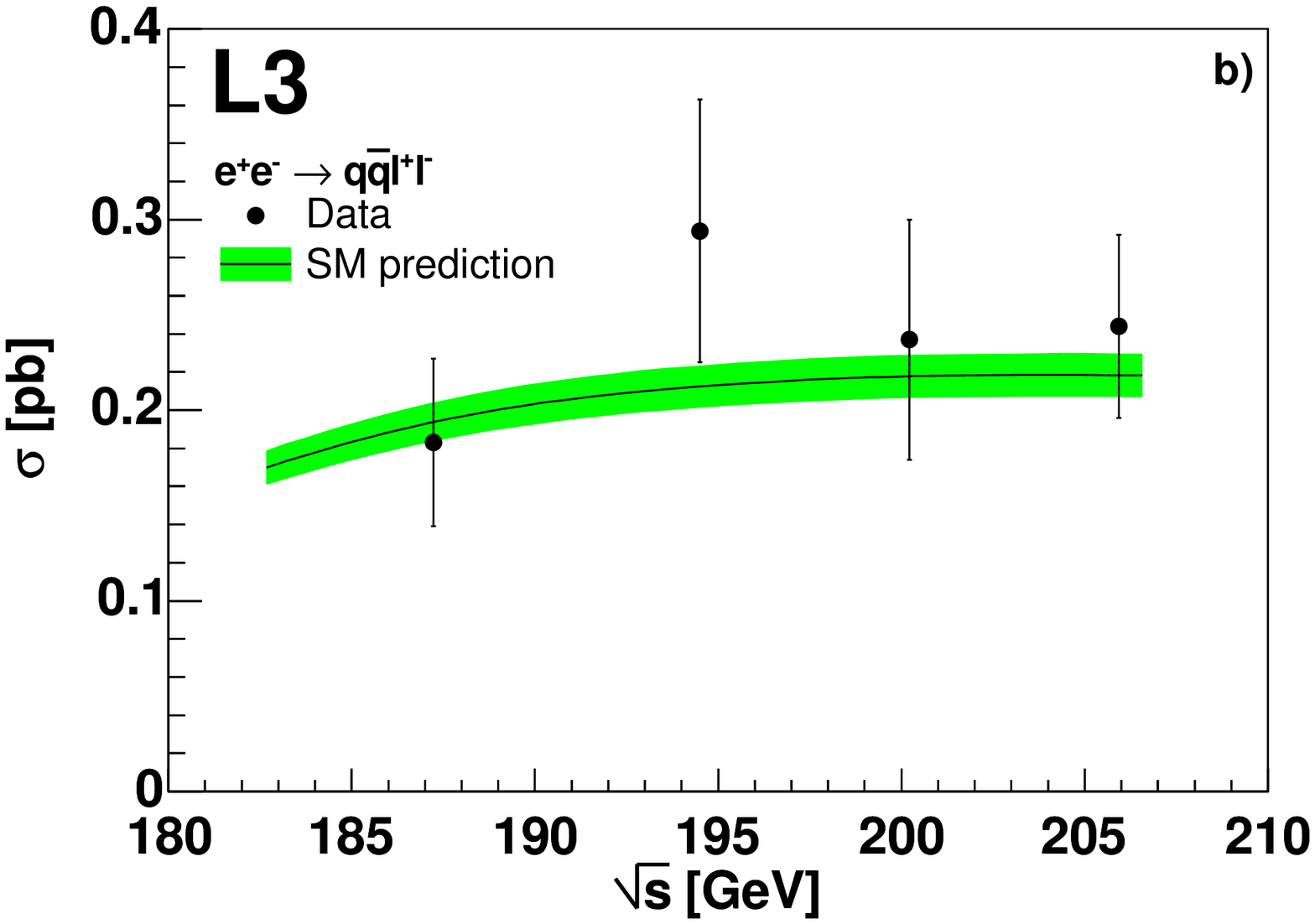}\\
    \includegraphics*[width=0.6\textwidth]{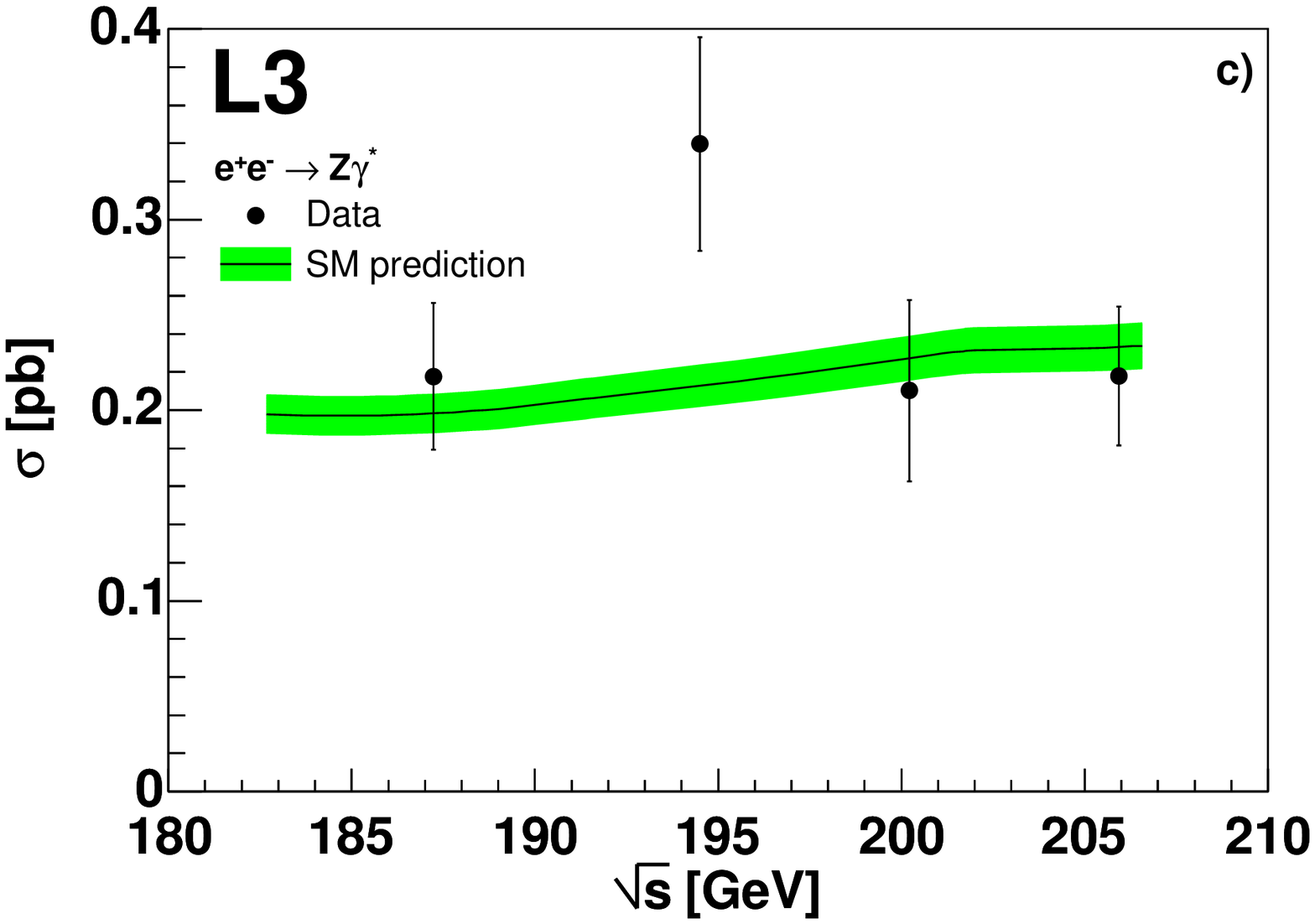}\\
    \end{tabular}
    \caption{Cross sections measured as a function of the
    centre-of-mass energies for the a) $\epem \ra \qqnn$, b) $\epem \ra
    \qqll$ and c) $\epem \ra{\rm Z} \gamma^* \ra f\bar{f}f'\bar{f}'$
    processes. The expectations from the GRC4F Monte Carlo are also
    shown, with an uncertainty of $\pm5\%$. Both the data and the
    predictions for the $\epem \ra \qqll$ process refer to the sum of
    the electron and muon final states.}
    \label{fig:6}
  \end{center}
\end{figure}

The high energies and high luminosity achieved by LEP have allowed
detailed studies of four-fermion production. Previous L3 studies
involving single and pair production of W and Z bosons are
complemented with a study of events from inclusive four-fermion
neutral-current production, $\epem \ra f\bar{f}f'\bar{f}'$, and from
the $\epem \ra{\rm Z} \gamma^* \ra f\bar{f}f'\bar{f}'$ process. Four
different final states are considered. Their cross sections are
measured and found to agree with the Standard Model predictions, as
shown in Table~\ref{tab:2}, Table~\ref{tab:4} and Figure~\ref{fig:6} for the
$\epem \ra \qqnn$, $\epem \ra \qqll$ and $\epem \ra{\rm Z} \gamma^*
\ra f\bar{f}f'\bar{f}'$ processes. The combined statistical and
systematic accuracy of the $\epem \ra{\rm Z} \gamma^* \ra
f\bar{f}f'\bar{f}'$ cross section measurement is 15\%. No evidence is
found for a new neutral hadronic-decaying heavy particle produced in
four-fermion events.

%
%
\bibliographystyle{thisL3stylem}

%
%
\newpage
\typeout{   }     
\typeout{Using author list for paper 287 -  }
\typeout{$Modified: Jul 15 2001 by smele $}
\typeout{!!!!  This should only be used with document option a4p!!!!}
\typeout{   }
%
%
%
%
%
%

\newcount\tutecount  \tutecount=0
\def\tutenum#1{\global\advance\tutecount by 1 \xdef#1{\the\tutecount}}
\def\tute#1{$^{#1}$}
\tutenum\aachen            
\tutenum\nikhef            
\tutenum\mich              
\tutenum\lapp              
\tutenum\basel             
\tutenum\lsu               
\tutenum\beijing           
\tutenum\bologna           
\tutenum\tata              
\tutenum\ne                
\tutenum\bucharest         
\tutenum\budapest          
\tutenum\mit               
\tutenum\panjab            
\tutenum\debrecen          
\tutenum\dublin            
\tutenum\florence          
\tutenum\cern              
\tutenum\wl                
\tutenum\geneva            
\tutenum\hamburg           
\tutenum\hefei             
\tutenum\lausanne          
\tutenum\lyon              
\tutenum\madrid            
\tutenum\florida           
\tutenum\milan             
\tutenum\moscow            
\tutenum\naples            
\tutenum\cyprus            
\tutenum\nymegen           
\tutenum\caltech           
\tutenum\perugia           
\tutenum\peters            
\tutenum\cmu               
\tutenum\potenza           
\tutenum\prince            
\tutenum\riverside         
\tutenum\rome              
\tutenum\salerno           
\tutenum\ucsd              
\tutenum\sofia             
\tutenum\korea             
\tutenum\taiwan            
\tutenum\tsinghua          
\tutenum\purdue            
\tutenum\psinst            
\tutenum\zeuthen           
\tutenum\eth               

{
\parskip=0pt
\noindent
{\bf The L3 Collaboration:}
\ifx\selectfont\undefined
 \baselineskip=10.8pt
 \baselineskip\baselinestretch\baselineskip
 \normalbaselineskip\baselineskip
 \ixpt
\else
 \fontsize{9}{10.8pt}\selectfont
\fi
\medskip
\tolerance=10000
\hbadness=5000
\raggedright
\hsize=162truemm\hoffset=0mm
\def\r{\rlap,}
\noindent

P.Achard\r\tute\geneva\ 
O.Adriani\r\tute{\florence}\ 
M.Aguilar-Benitez\r\tute\madrid\ 
J.Alcaraz\r\tute{\madrid}\ 
G.Alemanni\r\tute\lausanne\
J.Allaby\r\tute\cern\
A.Aloisio\r\tute\naples\ 
M.G.Alviggi\r\tute\naples\
H.Anderhub\r\tute\eth\ 
V.P.Andreev\r\tute{\lsu,\peters}\
% ...
\newpage
\begin{list}{A}{\itemsep=0pt plus 0pt minus 0pt\parsep=0pt plus 0pt minus 0pt
                \topsep=0pt plus 0pt minus 0pt}
\item[\aachen]
 III. Physikalisches Institut, RWTH, D-52056 Aachen, Germany$^{\S}$
\item[\nikhef] National Institute for High Energy Physics, NIKHEF, 
     and University of Amsterdam, NL-1009 DB Amsterdam, The Netherlands
\item[\mich] University of Michigan, Ann Arbor, MI 48109, USA
\item[\lapp] Laboratoire d'Annecy-le-Vieux de Physique des Particules, 
     LAPP,IN2P3-CNRS, BP 110, F-74941 Annecy-le-Vieux CEDEX, France
\item[\basel] Institute of Physics, University of Basel, CH-4056 Basel,
     Switzerland
\item[\lsu] Louisiana State University, Baton Rouge, LA 70803, USA
\item[\beijing] Institute of High Energy Physics, IHEP, 
  100039 Beijing, China$^{\triangle}$ 
\item[\bologna] University of Bologna and INFN-Sezione di Bologna, 
     I-40126 Bologna, Italy
\item[\tata] Tata Institute of Fundamental Research, Mumbai (Bombay) 400 005, India
\item[\ne] Northeastern University, Boston, MA 02115, USA
\item[\bucharest] Institute of Atomic Physics and University of Bucharest,
     R-76900 Bucharest, Romania
\item[\budapest] Central Research Institute for Physics of the 
     Hungarian Academy of Sciences, H-1525 Budapest 114, Hungary$^{\ddag}$
\item[\mit] Massachusetts Institute of Technology, Cambridge, MA 02139, USA
\item[\panjab] Panjab University, Chandigarh 160 014, India
\item[\debrecen] KLTE-ATOMKI, H-4010 Debrecen, Hungary$^\P$
\item[\dublin] Department of Experimental Physics,
  University College Dublin, Belfield, Dublin 4, Ireland
\item[\florence] INFN Sezione di Firenze and University of Florence, 
     I-50125 Florence, Italy
\item[\cern] European Laboratory for Particle Physics, CERN, 
     CH-1211 Geneva 23, Switzerland
\item[\wl] World Laboratory, FBLJA  Project, CH-1211 Geneva 23, Switzerland
\item[\geneva] University of Geneva, CH-1211 Geneva 4, Switzerland
\item[\hamburg] University of Hamburg, D-22761 Hamburg, Germany
\item[\hefei] Chinese University of Science and Technology, USTC,
      Hefei, Anhui 230 029, China$^{\triangle}$
\item[\lausanne] University of Lausanne, CH-1015 Lausanne, Switzerland
\item[\lyon] Institut de Physique Nucl\'eaire de Lyon, 
     IN2P3-CNRS,Universit\'e Claude Bernard, 
     F-69622 Villeurbanne, France
\item[\madrid] Centro de Investigaciones Energ{\'e}ticas, 
     Medioambientales y Tecnol\'ogicas, CIEMAT, E-28040 Madrid,
     Spain${\flat}$ 
\item[\florida] Florida Institute of Technology, Melbourne, FL 32901, USA
\item[\milan] INFN-Sezione di Milano, I-20133 Milan, Italy
\item[\moscow] Institute of Theoretical and Experimental Physics, ITEP, 
     Moscow, Russia
\item[\naples] INFN-Sezione di Napoli and University of Naples, 
     I-80125 Naples, Italy
\item[\cyprus] Department of Physics, University of Cyprus,
     Nicosia, Cyprus
\item[\nymegen] Radboud University and NIKHEF, 
     NL-6525 ED Nijmegen, The Netherlands
\item[\caltech] California Institute of Technology, Pasadena, CA 91125, USA
\item[\perugia] INFN-Sezione di Perugia and Universit\`a Degli 
     Studi di Perugia, I-06100 Perugia, Italy   
\item[\peters] Nuclear Physics Institute, St. Petersburg, Russia
\item[\cmu] Carnegie Mellon University, Pittsburgh, PA 15213, USA
\item[\potenza] INFN-Sezione di Napoli and University of Potenza, 
     I-85100 Potenza, Italy
\item[\prince] Princeton University, Princeton, NJ 08544, USA
\item[\riverside] University of Californa, Riverside, CA 92521, USA
\item[\rome] INFN-Sezione di Roma and University of Rome, ``La Sapienza",
     I-00185 Rome, Italy
\item[\salerno] University and INFN, Salerno, I-84100 Salerno, Italy
\item[\ucsd] University of California, San Diego, CA 92093, USA
\item[\sofia] Bulgarian Academy of Sciences, Central Lab.~of 
     Mechatronics and Instrumentation, BU-1113 Sofia, Bulgaria
\item[\korea]  The Center for High Energy Physics, 
     Kyungpook National University, 702-701 Taegu, Republic of Korea
\item[\taiwan] National Central University, Chung-Li, Taiwan, China
\item[\tsinghua] Department of Physics, National Tsing Hua University,
      Taiwan, China
\item[\purdue] Purdue University, West Lafayette, IN 47907, USA
\item[\psinst] Paul Scherrer Institut, PSI, CH-5232 Villigen, Switzerland
\item[\zeuthen] DESY, D-15738 Zeuthen, Germany
\item[\eth] Eidgen\"ossische Technische Hochschule, ETH Z\"urich,
     CH-8093 Z\"urich, Switzerland
\item[\S]  Supported by the German Bundesministerium 
        f\"ur Bildung, Wissenschaft, Forschung und Technologie.
\item[\ddag] Supported by the Hungarian OTKA fund under contract
numbers T019181, F023259 and T037350.
\item[\P] Also supported by the Hungarian OTKA fund under contract
  number T026178.
\item[$\flat$] Supported also by the Comisi\'on Interministerial de Ciencia y 
        Tecnolog{\'\i}a.
\item[$\sharp$] Also supported by CONICET and Universidad Nacional de La Plata,
        CC 67, 1900 La Plata, Argentina.
\item[$\triangle$] Supported by the National Natural Science
  Foundation of China.
\end{list}
}
\vfill


\newpage

%
%

\begin{table}
\begin{center}
\begin{tabular}{|c|*8{r}|r|}
\hline
$\sqrt{s}$~[Ge\kern -0.1em V]      & 182.7  & 188.6  & 191.6 & 195.5  & 199.5 & 201.7 & 205.1 & 206.8 & 196.6\\
$\mathcal{L}$~[pb$^{-1}$] & 55.4   & 176.8  & 29.7  &  83.7  &  82.7 &  37.1 &  69.1 & 138.9 & 673.4\\
\hline
\end{tabular}
\end{center}
\caption{Centre-of-mass energies and integrated luminosities for the
  different data-taking periods. The last column gives the
  luminosity-averaged centre-of-mass energy and the total integrated
  luminosity.}
\label{tab:1}
\end{table}

\begin{table}
\begin{center}
\begin{tabular}{|l|*6{c|}}
\hline
 \rule{0pt}{12pt}$\rm\epem\ra$ & $N_{\rm Data}$ & $N_{\rm Sign}^{\rm MC}$ & $N_{\rm Back}^{\rm MC}$ & \multicolumn{1}{c|}{$\varepsilon$}
& \multicolumn{1}{c|}{$\sigma_{4f}$~[pb]}  & $\sigma_{4f}^{\rm Th}$~[pb] \\ 
\hline
 \rule{0pt}{12pt}\qqnn & 198 & 73.2 & 125.8 & 38.1 \% & $0.278\pm 0.052\pm 0.021$  & 0.282 \\
 \rule{0pt}{12pt}\qqee & 109 & 60.4 & \phantom{0}37.8  & 59.1 \% & $0.156\pm 0.022\pm 0.006$  & 0.127 \\
 \rule{0pt}{12pt}\qqmm & \phantom{0}38  & 30.8 & \phantom{00}9.4   & 52.8 \% & $0.073\pm 0.016\pm 0.003$  & 0.082 \\
 \rule{0pt}{12pt}\llnn & \phantom{0}17  & \phantom{0}7.0  & \phantom{00}7.4   & 28.7 \% & $0.045\pm 0.022\pm 0.004$  & 0.036 \\
 \rule{0pt}{12pt}\llll & \phantom{0}25  & 14.8 & \phantom{00}9.9   & 39.9 \% & $0.058\pm 0.018\pm 0.004$  & 0.054 \\
\hline
\end{tabular} 
\end{center}
\caption{Numbers of  events observed for the $\epem\ra
  f\bar{f}f'\bar{f}'$ selections, $N_{\rm Data}$, compared to the Monte Carlo
  predictions for signal, $N_{\rm Sign}^{\rm MC}$, and background,
  $N_{\rm Back}^{\rm MC}$. The selection efficiencies,
  $\varepsilon$, are also given together with the measured cross
  sections, $\sigma_{4f}$ and the expectations from the GRC4F Monte
  Carlo, $\sigma_{4f}^{\rm Th}$. They refer to the luminosity-weighted
  average of the cross sections for each value of $\sqrt{s}$ in
  Table~2, corresponding to an average centre-of-mass energy
  $<\sqrt{s}>=196.6\GeV$. The first uncertainty on $\sigma_{4f}$ is
  statistical and the second systematic.}
\label{tab:2}
\end{table}

\begin{table}
\begin{center}
\begin{tabular}{|l|*6{c|}}
\hline
\rule{0pt}{12pt}$\rm\epem\ra\Zg\ra$ & $N_{\rm Data}$ & $N_{\rm Sign}^{\rm MC}$ & $N_{\rm Back}^{\rm MC}$ & \multicolumn{1}{c|}{$\varepsilon$}
& \multicolumn{1}{c|}{$\sigma_{\Zg}$~[pb]}  & $\sigma_{\Zg}^{\rm Th}$~[pb] \\ 
\hline
\rule{0pt}{12pt} \qqnn & 198 & 17.9 & 181.1 & 31.7 \% & $0.072\pm 0.044\pm 0.017$ & 0.083 \\
\rule{0pt}{12pt} \qqee & 109 & 23.5 & \phantom{0}74.7  & 58.8 \% & $0.100\pm 0.023\pm 0.007$ & 0.059 \\
\rule{0pt}{12pt} \qqmm & \phantom{0}38  & 14.0 & \phantom{0}26.3  & 49.2 \% & $0.040\pm 0.017\pm 0.004$ & 0.042 \\  
\rule{0pt}{12pt} \llnn & \phantom{0}17  & \phantom{0}3.2  & \phantom{0}11.3  & 27.5 \% & $0.039\pm 0.020\pm 0.004$ & 0.017 \\
\rule{0pt}{12pt} \llll & \phantom{0}25  & \phantom{0}5.3  & \phantom{0}19.5  & 45.7 \% & $0.019\pm 0.015\pm 0.004$ & 0.017 \\
\hline
\rule{0pt}{12pt} $ f\bar{f}f'\bar{f}'$& 387 & 63.9 & 312.6 & 43.2 \% &
$0.288\pm 0.052\pm 0.031 $ & 0.218 \\
\hline
\end{tabular}
\end{center}
\caption{Numbers of  events observed for the $\epem\ra{\rm Z} \gamma^*\ra
  f\bar{f}f'\bar{f}'$ selections, $N_{\rm Data}$, compared to the Monte Carlo
  predictions for signal, $N_{\rm Sign}^{\rm MC}$, and background,
  $N_{\rm Back}^{\rm MC}$. The selection efficiencies,
  $\varepsilon$, are also given together with the measured cross
  sections, $\sigma_{\Zg}$ and the expectations from the GRC4F Monte
  Carlo, $\sigma_{\Zg}^{\rm Th}$. They refer to the luminosity-weighted
  average of the cross sections for each value of $\sqrt{s}$ in
  Table~2, corresponding to an average centre-of-mass energy
  $<\sqrt{s}>=196.6\GeV$. The first uncertainty on $\sigma_{\Zg}$ is
  statistical and the second systematic.}
\label{tab:4}
\end{table}

\begin{table}
\begin{center}
\begin{tabular}{|l|*{10}{r|}c|}
\cline{2-12}
 \multicolumn{1}{c|}{}  & \multicolumn{2}{c|}{\qqee} & \multicolumn{2}{c|}{\qqmm}   & \multicolumn{2}{c|}{\qqnn}    & \multicolumn{2}{c|}{\llnn}     & \multicolumn{2}{c|}{\llll}     & \multicolumn{1}{c|}{Comb.} \\ 
\hline
Source                   & 4$f$   & \Zg  & 4$f$    & \Zg   & 4$f$   & \Zg & 4$f$   & \Zg  & 4$f$   & \Zg    & \Zg \\
\hline
Energy scale             & 0.7  & 0.7   & 0.7  & 0.7   & 3.6  & 3.6 & ---  &  --- & ---  &  --- &  \phantom{0}1.1 \\
Lepton identification    & 1.9  & 1.9   & 1.9  & 1.9   & ---  & --- & 0.8  & 0.8  & 0.7  &  0.7 &  \phantom{0}1.4 \\
Cut variation            & 2.2  & 2.2   & 2.2  & 2.2   & 3.6  & 3.6 & 4.8  & 4.8  & 4.6  &  4.6 &  \phantom{0}2.7 \\
MC statistics (sign.)    & 0.6  & 0.6   & 0.5  & 0.2   & 0.2  & 0.3 & 0.4  & 0.7  & 0.4  &  0.5 &  \phantom{0}0.5 \\
MC statistics (back.)    & 2.2  & 2.2   & 3.4  & 3.5   & 0.7  & 0.7 &$<$0.1& 0.7  &$<$0.1&  0.3 &  \phantom{0}2.4 \\
Background normalisation & 1.5  & 6.8   & 0.5  & 8.4   & 5.6  &22.4 & 8.2  & 8.6  & 5.7  & 19.2 &  \phantom{0}9.9 \\
\hline											      
Total                    & 4.0  & 7.8   & 4.6  & 9.6   & 7.6  &23.0 & 9.5  & 9.9  & 7.4  & 19.8 & 10.7 \\
\hline
\end{tabular}
\end{center}
\caption{Sources and effects of systematic uncertainties. Values are
  given as the percentual variation on the measured cross sections of the
  $\epem\ra f\bar{f}f'\bar{f}'$ ($4f$) and $\epem \ra{\rm Z} \gamma^* \ra f\bar{f}f'\bar{f}'$ 
  (\Zg) processes. The last column refers to the combination of the
  channels used to measure the  $\epem \ra{\rm Z} \gamma^* \ra
  f\bar{f}f'\bar{f}'$ cross section. The total systematic uncertainty
  is the sum in quadrature of the different contributions.}
\label{tab:3}
\end{table}

%
%

\end{document}